\documentclass[12pt]{iopart}

\usepackage{epsfig}
\usepackage{ifpdf}
\usepackage{graphicx}

\usepackage{amssymb,lineno,mathbbol,upgreek}

\usepackage[tight]{subfigure}

\pdfoutput=1

\usepackage{bbm}
\usepackage{slashed}
\usepackage{calligra}
\DeclareMathAlphabet{\mathcalligra}{T1}{calligra}{m}{n}
\DeclareFontShape{T1}{calligra}{m}{n}{<->s*[2.2]callig15}{}

\usepackage{calligra}
\DeclareMathAlphabet{\mathcalligra}{T1}{calligra}{m}{n}
\DeclareFontShape{T1}{calligra}{m}{n}{<->s*[2.2]callig15}{}
\usepackage{enumitem}

\begin{document}

\title[Relativistic solitons in armchair nanoribbon optical lattices]{The nonlinear Dirac equation in Bose-Einstein condensates: I. Relativistic solitons in armchair nanoribbon optical lattices}

\author{L H Haddad$^1$, C M Weaver$^1$ and Lincoln D Carr$^{1,2}$}
\address{$^1$Department of Physics, Colorado School of Mines, Golden, CO 80401,USA  \\$^2$Physikalisches Institut, Universit\"at Heidelberg, D-69120 Heidelberg, Germany}

\ead{\mailto{laith.haddad@gmail.com}, \mailto{lcarr@mines.edu}}

\begin{abstract}
We present a thorough analysis of soliton solutions to the quasi-one-dimensional nonlinear Dirac equation (NLDE) for a Bose-Einstein condensate in a honeycomb lattice with armchair geometry. Our NLDE corresponds to a quasi-one-dimensional reduction of the honeycomb lattice along the zigzag direction, in direct analogy to graphene nanoribbons. Excitations in the remaining large direction of the lattice correspond to the linear subbands in the armchair nanoribbon spectrum. Analytical as well as numerical soliton Dirac spinor solutions are obtained. We analyze the solution space of the quasi-one-dimensional NLDE by finding fixed points, delineating the various regions in solution space, and through an invariance relation which we obtain as a first integral of the NLDE. We obtain spatially oscillating multi-soliton solutions as well as asymptotically flat single soliton solutions using five different methods: by direct integration; an invariance relation; parametric transformation; a series expansion; and by numerical shooting. By tuning the ratio of the chemical potential to the nonlinearity for a fixed value of the energy-momentum tensor, we can obtain both bright and dark solitons over a nonzero density background.  
\end{abstract}

\submitto{\NJP}
\maketitle

\section{Introduction}
\label{introduction}

The nonlinear Dirac equation (NLDE) appears in a variety of physical settings, typically as classical field equations for relativistic interacting fermions~\cite{Lee1975,Cooper2010}. In fact, the (1+1)-dimensional NLDE with scalar-scalar or vector-vector interaction is the prototypical effective model for interacting fermions, and has been the subject of much analysis over the past decades~\cite{fushchich89,toyama94,liangzhong00,Parwani2008,cazenave86,estebanMJ1995}. Recently, analytical solutions of the massive NLDE were obtained for the case of Kerr nonlinearity~\cite{Alkhawaja2014}. Dirac-like spin-orbit couplings for interacting cold atoms have also been investigated, simulating some features of quark confinement~\cite{santos2009}. Moreover, solitons appear in systems with Dirac points such as quasi-one-dimensional (quasi-1D) nonlinear optical structures~\cite{Christodoulides1988,Segev2000,Efremidis2003,Peleg2007,Bahat2008,ablowitz2009,Segev2010,Kartashov2011,Kartashov2013,Guo2013,Plotnik2014}, acoustic physics~\cite{TorSan12}, and electron propagation in graphene~\cite{Geim2009,katsnelson2006,Semenoff1984,Mele1984}. In all of these cases the combination of Dirac kinetic term and nonlinearity leads to a plethora of solitary wave solutions whose properties depend on the particular form of the interaction term~\cite{Parwani2008,Lee1990}. We note that the (1+1)-dimensional nonlinear Dirac equation has also been obtained starting from the nonlinear Schr\"odinger equation in a periodic potential using an asymptotic multi-scale expansion method~\cite{Pelinovsky2011}. Our own recent work has placed the NLDE in the context of a Bose-Einstein condensate (BEC)~\cite{haddad2009}. Significantly, our particular form of the NLDE has opened up research in other fields of physics~\cite{Park2009,Block2010,Dellar2011,Ablowitz2010,CB2010,Chen2011,Kapit2011,Zhang2009,Gupta2010}. For the NLDE in a BEC, the relativistic structure arises naturally as bosons propagate in a shallow periodic honeycomb lattice potential, and yields a rich soliton landscape which we explore in detail in this Article.

\begin{figure}[h]
\centering
 \subfigure{
\label{fig:ex3-a}
\includegraphics[width=.925\textwidth]{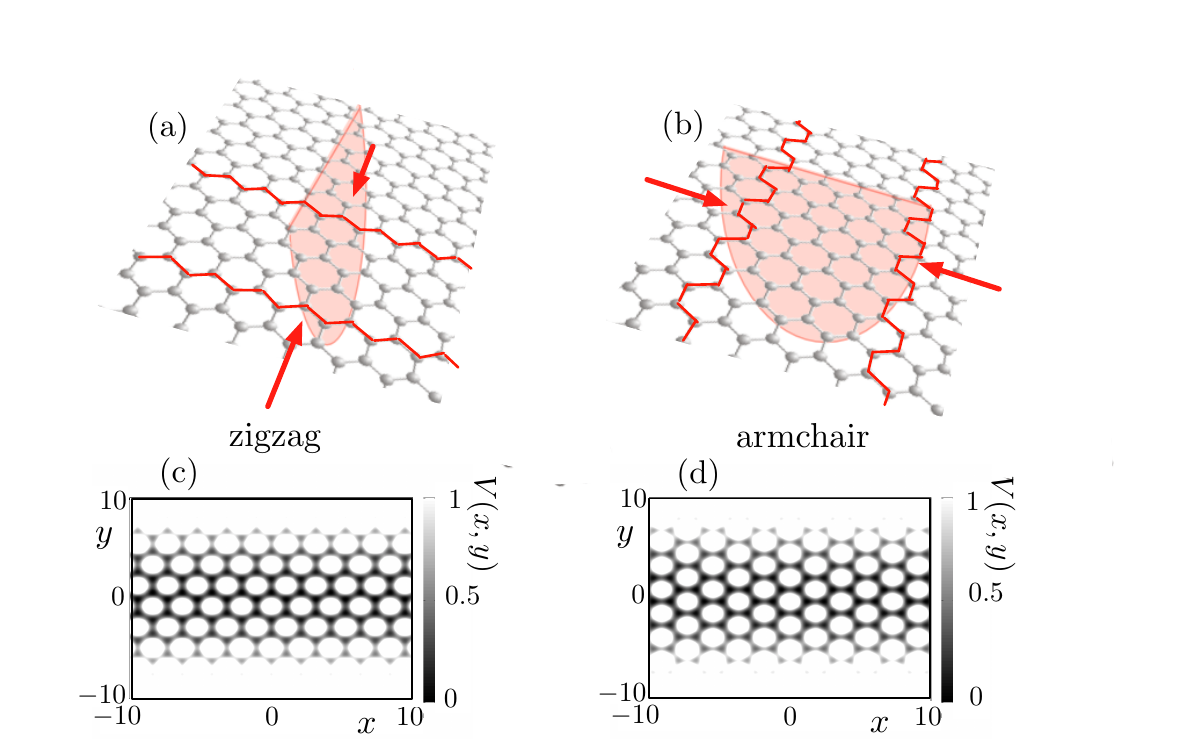} }\\
\caption[]{\emph{Quasi-one-dimensional reduction of a BEC in a honeycomb optical lattice}. (a)-(b) Harmonic confining potential parallel to the plane of the lattice produces either the zigzag or the armchair pattern, depending on its orientation in the plane. The red lines in the plane indicate the nanoribbon boundaries. (c)-(d) Harmonic trap + honeycomb lattice potentials.} 
\label{OneD}
\end{figure}

For graphene nanoribbons the single-particle spectrum associated with the zigzag direction contains edge states as well as states confined to the interior of the ribbon, but none of these states have a linear degenerate band with Dirac-like dispersion. In contrast, for certain nanoribbon widths the spectrum of the armchair ribbon does contain Dirac points~\cite{Dresselhaus1996,Brey2006,Geim2009}. The spectra and single-particle states for armchair and zigzag ribbons have been obtained using both a tight-binding Schr\"odinger calculation~\cite{Sigrist2009,Koichi1996}, and by starting from the Dirac Hamiltonian for the long-wavelength limit of the 2D lattice~\cite{Brey2006,Geim2009}. Both methods agree quite well in the intermediate to large ribbon width range, i.e., for $N \gtrsim 10$ where $N$ is the number of zigzag or armchair lines. However, Dirac points only occur in the spectrum for the armchair case and here only for particular values of $N$. By imposing quantization conditions at the edges of the ribbon one obtains a restriction on $N$ to integer multiples of 3. Thus, for fixed values of the lattice constant one identifies specific ribbon widths that possess the dispersion of interest to us. This is similar to the case of graphene nanotubes where different $(m,n)$ combinations have different properties due to different 1D cuts through the Dirac cone (some are semiconducting, others semimetals). In practice, the quasi-1D NLDE is obtained by isolating the armchair direction in the full two-dimensional (2D) honeycomb lattice theory~\cite{haddad2009}. This is accomplished by starting with the 2D lattice. One then increases the trap potential in one of the planar directions until a desired effective width is obtained. The Dirac theory of the full 2D lattice is modified by introducing armchair boundary conditions for the short direction. This leaves a translationally invariant theory in the long direction described by either a massive or massless 1D NLDE (armchair NLDE) determined by the particular subband~\cite{Sigrist2009,Brey2006,Enoki2010}. In our case, we focus on confinement in the zigzag direction, i.e., transverse to the armchair pattern of the 2D lattice. A schematic of the harmonic magnetic trap, interfering lasers, and BEC required to realize our set up is shown in Fig.~\ref{OneD}. Details of the experimental construction can be found in~\cite{Haddad2012}.

There are two equivalent forms of the quasi-1D reduction of the NLDE corresponding to a real or complex projection of the Dirac operator in one spatial dimension. Consequently, spinor solutions associated with these two projections are related by a complex Pauli matrix rotation. To obtain soliton solutions we first integrate the armchair NLDE to obtain an invariance relation which describes solutions at fixed values of the diagonal spatial element of the energy-momentum, a quantity which may be positive or negative valued in relation to zero energy set at the Dirac point. The invariance relation provides a vantage point which offers insight into general solutions of the quasi-1D reduction of the NLDE. In particular, we find soliton solutions residing at the boundary between two oscillating solution regimes. We will refer to this boundary in parameter space as the \emph{soliton boundary}. The soliton boundary appears for a particular value of the ratio $\mu/U = (\mu/U)_\mathrm{SB}$, where $\mu$ is the chemical potential of the system and $U$ is the quasi-one-dimensional renormalized interaction. Tuning $\mu/U$ towards $(\mu/U)_\mathrm{SB}$ while maintaining the local particle density above some critical value, we encounter there a bright soliton, whereas a dark soliton is obtained for densities less than the critical value. Oscillating solutions away from the soliton boundary correspond to multiple dark or bright soliton and are not necessarily stable. However, the single solitons at the soliton boundary are robust objects, as we will explain in Sec.~\ref{GeneralSolutions}.

To better understand the two types of solitons, we consider the two-dimensional solution space that results from fixing the internal and overall phase of a Dirac two-spinor. The two types of solitons correspond to paths in solution space that interpolate between two fixed points and pass along either the small amplitude (dark soliton) or the large amplitude (bright soliton) side of a third fixed point. The two paths (and associated solitons) are topologically distinct. Our analysis centers on solutions of the armchair NLDE with real Dirac operator but extension of our results to the complex form via the aforementioned Pauli rotation is straightforward. The work presented in this article is devoted to finding single and multi-soliton solutions of the NLDE. We have chosen to do this using several methods to emphasize the multiple lines of evidence for dark and bright solitons. 

The work that we present here is related to a number of parallel studies in condensed matter and cold atomic gases, among other contexts. The dimensional reduction of the quasi-2D honeycomb lattice to a quasi-1D lattice provides a novel way to study BECs. Another approach which has been proposed for simulating Dirac fermions using cold bosonic atoms relies on laser-induced spin-orbit coupling in a spinor BEC~\cite{santos2009}. The hyperfine structure provides the internal degrees of freedom needed to simulate spin while the additional lasers couple spinor states to the spatial degrees of freedom. We note that in our case both of these effects come from the lattice background and are therefore geometric in origin. It is the combination of nonlinearity and Dirac spin structure which allows for self-localization similar to chiral confinement in relativistic models such as the massive Thirring and Gross-Neveu models~\cite{santos2009,basar2008a,basar2008b,basar2011,Ward1988}. Our main interest is not in simulating Dirac fermions per se, but exploring a relativistic nonlinear system in the highly tunable and controllable context of BECs, where effective relativistic velocities are 10 orders of magnitude slower than the speed of light~\cite{Haddad2012}.

This article is organized as follows. In Sec.~\ref{nlde}, we provide an introduction to the NLDE and discuss the key physical parameters. In Sec.~\ref{quasioned}, we explain how the NLDE armchair geometry is realized starting from the 2D honeycomb lattice. This step is essential in order to establish an experimental foundation for the rest of this paper. In Sec.~\ref{SolutionProperties}, we determine general properties of the NLDE solution space. It is important to note that we treat only stationary solutions in this article; the question of dynamics is retained as a subject of future work. Focusing on the time-independent armchair NLDE, we find all the fixed points and regions of solution space according to the character of the associated direction fields, i.e., the vectors formed from the spatial first derivatives of the two-spinor components. We also derive the main invariance relation governing the NLDE which leads to explicit soliton solutions. In Sec.~\ref{GeneralSolutions}, we use the insight obtained by our study of fixed points and invariance relations to map out the phase diagram for NLDE solutions. In Sec.~\ref{AnalyticalSolutions}, we solve the NLDE analytically using a trigonometric ansatz, through detailed analysis of our invariance relation, using a parametric transformation and by a power series expansion. In Sec.~\ref{NumericalSolitons}, we obtain solitons using a numerical shooting method. Finally, in Sec.~\ref{Conclusion} we conclude.

\section{The nonlinear Dirac equation}
\label{nlde}

In this section we introduce the NLDE, a nonlinear extension of the massless Dirac equation, and discuss the experimentally relevant physical parameters. The NLDE for two inequivalent Dirac points describes the dynamics of a Dirac four-spinor of the form $\Psi \equiv  \left( \Psi_+ , \, \Psi_- \right)^T$, with the upper ($+$) and lower ($-$) two-spinors relating to opposite ${\bf K}$ and ${\bf K}^\prime$ points of the honeycomb lattice (see Refs.~\cite{Semenoff1984,Mele1984,haddad2009,haddad2011}). We remind the reader that Dirac points are locations in the single particle spectrum where the upper and lower energy bands become degenerate (the energy bands cross) with a linear structure, i.e., $\mathrm{E}(\mathrm{k}) \approx  \hbar v \mathrm{k}$, a consequence of the underlying symmetry of the honeycomb lattice. For graphene the proportionality constant $v$ is just the Fermi velocity $v_F$. In BECs $v$ is the quasi-particle group velocity $c_l$, and required to be less than the speed of sound in order to satisfy the Landau criterion. Note that in both cases the Dirac point is a kinetic single-particle effect, where $v$ is determined by the microscopic physics and plays the role of an effective speed of light. In terms of the A and B sublattice wavefunctions, we have $\Psi_+ \equiv \left(  \psi_{A +} \, , \;   \psi_{B+} \right)^T$ and $\Psi_- \equiv \left(  \psi_{B - } \, , \;   \psi_{A-} \right)^T$. The full NLDE in this case is 
 \begin{eqnarray}
i \hbar    \gamma^\mu  \partial_\mu   {\Psi}  +  U_\mathrm{2D} \sum_{i = 1}^4 (\Psi^\dagger \mathrm{M}_i )( \Psi  \mathrm{M}_i)  \, \Psi = 0   \, .             \label{NLDE1}
\end{eqnarray} 
 The matrices $\gamma^\mu$ are the usual Dirac matrices and the interaction terms are encapsulated in the summation with the matrices $\mathrm{M}_i$ constructed to give the correct cubic nonlinearities, local to each spinor component~\cite{haddad2009}. Explicitly, the interaction matrices are 
  \begin{eqnarray}
   \mathrm{M}_1 = \left(
  \begin{array}{ c c c c }
       1   &  0  & 0 & 0 \\
       0    &    0 & 0 & 0  \\
        0    &    0 & 0 & 0  \\
         0    &    0 & 0 & 0 
  \end{array} \right)     \,    ,   \;\;\;    \mathrm{M}_2 = \left(
  \begin{array}{ c c c c }
       0  &   0 & 0 & 0   \\
       0   &  1 & 0 & 0 \\
        0   &  0 & 0 & 0 \\
        0   &  0 & 0 & 0 
  \end{array} \right)      \,         ,    \\
      \nonumber  \\
  \mathrm{M}_3 = \left(
  \begin{array}{ c c c c }
       0   &  0  & 0 & 0 \\
       0    &    0 & 0 & 0  \\
        0    &    0 & 1 & 0  \\
         0    &    0 & 0 & 0 
  \end{array} \right)     \,    ,   \;\;\;    \mathrm{M}_4 = \left(
  \begin{array}{ c c c c }
       0  &   0 & 0 & 0   \\
       0   & 0  & 0 & 0 \\
        0   &  0 & 0 & 0 \\
        0   &  0 & 0 & 1
  \end{array} \right)      \,     . 
\end{eqnarray}
In this simplest version of the NLDE for a BEC in a honeycomb lattice, the interactions do not couple different spinor components, which are only coupled through the kinetic term.$^{\footnotemark[1]}$ \footnotetext[1]{Note that were we to consider nearest-neighbor interaction terms in the quantum Hamiltonian, we would have such a coupling; however, here we make the usual tight-binding, lowest band approximation, as such intersite interaction terms are small -- see Ref.~\cite{Haddad2012}.}Thus, in full generality we can focus on the equations for a two-spinor in rectangular coordinates while omitting the Dirac point subscript
\begin{eqnarray}
 - i \hbar  c_l \left( \partial_x   -  i \partial_y \right) \psi_B +  U_\mathrm{2D} \left| \psi_A\right|^2 \psi_A  =  i \hbar \,  \partial_t \psi_A \, ,\label{eqn:CondPsi5} \\
  - i \hbar c_l \left( \partial_x   + i \partial_y \right) \psi_A +  U_\mathrm{2D} \left| \psi_B \right|^2 \psi_B  =   i \hbar \,  \partial_t \psi_B  \label{eqn:CondPsi6}  \,  , 
\end{eqnarray}
with the full solution expressed as a linear combination of solutions from each Dirac point. Note the presence of the effective speed of light, $c_l$, and interaction strength, $U_\mathrm{2D}$.

Equations~(\ref{eqn:CondPsi5})-(\ref{eqn:CondPsi6}) allow for quasi-one-dimensional (quasi-1D) solutions by confining the BEC in one of the planar directions. From here on we will assume confinement in the $y$-direction. The experimental construction is shown in Fig.~\ref{OneD} where the BEC resides within a weak magnetic harmonic trap and an optical potential created by three laser beams offset by relative angles of $120^\mathrm{o}$ in the $x$ $y$ plane. Confinement in the $y$-direction produces an optical lattice with ribbon geometry wherein  $\psi_A$ and $\psi_B$ are effectively functions only of $x$, for energies small compared to the characteristic energy associated with the width. A full derivation of the dimensional reduction from the 2D NLDE to the quasi-1D form is presented in Sec.~\ref{quasioned}. The stationary states of interest are then obtained by taking the time-dependence to be the usual exponential factor with the chemical potential $\mu$ as the frequency: $\psi_A(x, t) = \,  \mathrm{exp}(-i \mu t/\hbar) f_A(x)$ and  $\psi_B(x, t) =  \,  \mathrm{exp}(-i \mu t/\hbar) f_B(x)$. Equations~(\ref{eqn:CondPsi5})-(\ref{eqn:CondPsi6}) become 
\begin{eqnarray}
       - i  \hbar  c_l  \, \partial_x f_B(x)  +  U_\mathrm{1D} \,  |f_A(x)|^2 f_A(x)  =    \mu f_A(x)\label{eqn:CondPsi7}\, , \\
        -i  \hbar  c_l  \, \partial_x f_A(x)  +  U_\mathrm{1D}  \,  |f_B(x)|^2 f_B(x)  =    \mu f_B(x) \label{eqn:CondPsi8}  \, . 
\end{eqnarray}
\footnotetext[1]{Note that were we to consider nearest-neighbor interaction terms in the quantum Hamiltonian, we would have such a coupling; however, here we make the usual tight-binding, lowest band approximation, as such intersite interaction terms are small -- see Ref.~\cite{Haddad2012}.}The system, Eqs.~(\ref{eqn:CondPsi7})-(\ref{eqn:CondPsi8}), is the time-independent quasi-1D NLDE for the complex form of the Dirac operator. Notice that taking $f_A \to  i f_B$ and $f_B \to f_A$ converts Eqs.~(\ref{eqn:CondPsi7})-(\ref{eqn:CondPsi8}) to the form associated with the real Dirac operator, as can also be obtained by choosing the $y$-direction in Eqs~(\ref{eqn:CondPsi5})-(\ref{eqn:CondPsi6}). This transformation is equivalent to multiplication by a linear combination of Pauli matrices with an overall complex factor, $[(1 + i)/2] (\sigma_x - \sigma_y)$. It is natural to  retain the $x$ notation when discussing either form of the NLDE; thus we write the real form as 
\begin{eqnarray}
  - \hbar  c_l  \, \partial_x f_B(x)  +  U_\mathrm{1D} \,  |f_A(x)|^2 f_A(x) =   \mu f_A(x)  \label{eqn:CondPsi9}  \, , \\
    \hbar  c_l  \, \partial_x f_A(x)  +  U_\mathrm{1D} \,  |f_B(x)|^2 f_B(x)  =   \mu f_B(x)  \label{eqn:CondPsi10}  \, . 
\end{eqnarray}
In Sec.~\ref{quasioned}, we will see that the interaction $U_\mathrm{1D}$ in Eqs.~(\ref{eqn:CondPsi7})-(\ref{eqn:CondPsi10}) is the renormalized version of the corresponding 2D interaction $U_\mathrm{2D}$ in Eqs.~(\ref{eqn:CondPsi5})-(\ref{eqn:CondPsi6}).

Finally, we provide a brief consideration of physical implementation of the NLDE in BECs, with more detailed treatment given in~\cite{Haddad2012}. The parameters which enter directly into the NLDE and will therefore appear in all of our solutions, are the effective speed of light $c_l = t_h a \sqrt{3}/2 \hbar$; and the atom-atom binary interaction strength, $U_\mathrm{2D} = L_z g \bar{n}^2 3 \sqrt{3} a^2/8$, where $U_\mathrm{2D}$ is the 2D optical lattice renormalized version of the usual interaction $g= 4 \pi \hbar^2 a_s/M$. Appearing in these definitions are the average particle density $\bar{n} = N/V$, the $s$-wave scattering length $a_s$, the vertical oscillator length $L_z$, the mass $M$ of the constituent atoms in the BEC, the lattice constant $a$, and the hopping energy $t_h$. For the hopping energy, we use a semiclassical estimate given by $t_h  \equiv 1.861 \left(V_0/E_R\right)^{1/4} E_R\, \mathrm{exp}\! \left( -1.582 \sqrt{V_0/E_R} \right)$~\cite{Lee:2009}, where $V_0$ and $E_R$ are the lattice potential depth and recoil energy. Typical practical values of key physical parameters in order to realize NLDE solitons in the quasi-one-dimensional regime (which we will discuss in detail in Sec.~\ref{quasioned}) are $T = 8.0 \,  \mathrm{nK} $, $\bar{n} =  1.5 \times 10^{9} \, \mathrm{cm}^{-3}$, $L_x \gg L_y  \gg L_z  =  3.0 \,  \mathrm{\mu m}$, $t_h = 4.31 \, \mathrm{nK} = 1.04 \, \mathrm{kHz}$, $V_0/E_R = 16$, $U_\mathrm{2D} =  0.391\,  \mathrm{pK} = 0.051 \, \mathrm{Hz}$, $a = 0.55 \,  \mathrm{\mu m}$, where $T$ is the temperature and $L_y$ is the transverse oscillator length parallel to the plane of the honeycomb lattice. For this typical parameter set we took the atomic mass $M$ to be that of $^{87}\mathrm{Rb}$, and the associated scattering length $a_s = 5.77 \, \mathrm{nm}$. A complete discussion of NLDE parameters and constraints can be found in~\cite{Haddad2012}.

\section{Quasi-one-dimensional reduction in the honeycomb lattice}
\label{quasioned}

The quasi-1D NLDE is physically realized by starting with the 2D honeycomb lattice, then adding harmonic confinement in the $y$-direction. For generality we will explain the construction of both the armchair and zigzag geometries. We then focus exclusively on the armchair case since here the spectrum contains the linear degenerate points crucial to our results. The potentials for the honeycomb lattice with harmonic confinement for the armchair and zigzag geometries are given by 
\begin{eqnarray}
\fl V_\mathrm{armchair}({ \bf r}) =  - \frac{\alpha E_0^2}{4} \, \left( 3 + 2 \, \mathrm{cos}\left[ (\mathrm{\bf k}_1 - \mathrm{\bf k}_2 ) \cdot {\bf r} \right] + 2 \, \mathrm{cos}\left[ (\mathrm{\bf k}_2 - \mathrm{\bf k}_3 ) \cdot {\bf r} \right] \right. \nonumber \\
\left.   + 2 \, \mathrm{cos}\left[ (\mathrm{\bf k}_1 - \mathrm{\bf k}_3 ) \cdot {\bf r} \right]  \right) + \frac{1}{2} M \omega_y^2 y^2 \, , \label{Varmchair} \\
\fl V_\mathrm{zigzag}( { \bf r}) =  - \frac{\alpha E_0^2}{4} \, \left( 3 + 2 \, \mathrm{cos}\left[ (\mathrm{\bf k}_2 - \mathrm{\bf k}_1 ) \cdot {\bf r} \right] + 2 \, \mathrm{cos}\left[ (\mathrm{\bf k}_1 - \mathrm{\bf k}_3 ) \cdot {\bf r} \right] \right.  \nonumber \\
\left.  +\,  2 \, \mathrm{cos}\left[ (\mathrm{\bf k}_2 - \mathrm{\bf k}_3 ) \cdot {\bf r} \right]  \right) + \frac{1}{2} M \omega_y^2 y^2 \, ,\label{Vzigzag}
\end{eqnarray}
where $\alpha$ is the polarizability of the atoms, $E_0$ is the electric field strength,  ${\bf r} = (x, y)$ is the planar  coordinate vector, $M$ is the atomic mass of the constituent bosons, and $\omega_y$ is the frequency of the harmonic potential which adds the additional confinement. The wave vectors $\mathrm{\bf k}_1$, $\mathrm{\bf k}_2$, and $\mathrm{\bf k}_3$ in Eqs.~(\ref{Varmchair})-(\ref{Vzigzag}) are defined as
\begin{eqnarray}
\mathrm{\bf k}_1  =  k \,  \hat{\mathrm{ \bf y}} \, , \\
\mathrm{\bf k}_2  =   \left( \frac{\sqrt{3}}{2} \hat{\mathrm{ \bf x}} -  \frac{1}{2} \hat{\mathrm{ \bf y}} \right) k \, , \\
\mathrm{\bf k}_3   =   \left( -\frac{ \sqrt{3}}{2} \hat{\mathrm{ \bf x}} -  \frac{1}{2} \hat{\mathrm{ \bf y}} \right) k\, . 
\end{eqnarray}
The different forms of $V_\mathrm{armchair}$ and $V_\mathrm{zigzag}$ in Eqs.~(\ref{Varmchair})-(\ref{Vzigzag}) reflect a uniform rotation of the beams by $90^\mathrm{o}$. The wavelength of the laser light that forms the lattice is $\lambda = 2 \pi/k$ which defines the distance between sites on a hexagonal sublattice as $a = 2 \lambda/3$. The underlying electric field which produces the potentials in Eqs.~(\ref{Varmchair})-(\ref{Vzigzag}) is the superposition of the fields from each beam and is given explicitly by 
\begin{eqnarray}
\mathrm{\bf E} = \frac{E_0}{2} \hat{ \mathrm{ \bf z} }\sum_{ i =1}^3 \mathrm{exp} \left[ i \left( \mathrm{ \bf k}_i  \cdot  {\bf r } + \omega_i t \right) \right]  \, , 
\end{eqnarray}
where $\omega_i = c |\mathrm{\bf k}_i|$ ($c =$ speed of light $\approx  2.99 \times 10^8\,  \mathrm{m} \cdot  \mathrm{s}^{-1}$) here must be distinguished from the trapping frequency, and $\hat{\mathrm{ \bf z}}$ indicates the polarization of the beams perpendicular to the plane of the lattice. In the AC stark effect relevant to ultracold atoms trapped in optical lattices, the lattice potential is then obtained by taking the average of the square of the electric field
\begin{eqnarray}
V = - \frac{1}{2} \alpha \overline{|\mathrm{\bf E}|^2} \, , 
\end{eqnarray}
where $V$ here can stand for either potential in Eqs.~(\ref{Varmchair})-(\ref{Vzigzag}). Plots of the armchair and zigzag potentials are shown in Fig.~\ref{OneD}(c)-(d). For the remainder of our work we focus exclusively on the armchair configuration.

We now solve Eqs.~(\ref{eqn:CondPsi5})-(\ref{eqn:CondPsi6}) in the presence of the armchair potential in Eq.~(\ref{Varmchair}). Stationary states are obtained as in the unconfined case but now with additional armchair boundary conditions in the $y$-direction for $\psi_A$ and $\psi_B$. Respecting translational invariance along the $x$-direction, separation of variables for the spinor functions gives $\psi_A(x, y, t)$$ = \,  \mathrm{exp}[ i (k_x x -   \mu t/\hbar  )] \, $$\eta(y) g_A(y)$ and  $\psi_B(x, y, t) $$=   \,  \mathrm{exp}[ i ( k_x x -  \mu t/\hbar  )] \, $$\eta(y) g_B(y)$, where $\eta$ is a real envelope function and $g_{A(B)}$ are superpositions of left and right plane wave excitations along the width of the ribbon. Equations~(\ref{eqn:CondPsi5})-(\ref{eqn:CondPsi6}) become 
\begin{eqnarray}
      &&\hspace{-5pc}  \hbar  c_l   \left(  k_x   -  \partial_y \right)    g_B     -   \hbar c_l   g_B \frac{ \partial_y \eta}{ \eta}     +             U_\mathrm{2D} \,   \eta^2  | g_A|^2 g_A    +  \frac{1}{2} M \omega_y^2  y^2    \, g_A =    \mu  \, g_A   \label{CondPsi7}\, , \\
      &&\hspace{-5pc}  \hbar  c_l   \left(  k_x   +   \partial_y \right)    g_A     +   \hbar c_l   g_A \frac{ \partial_y \eta}{ \eta}     +             U_\mathrm{2D} \,    \eta^2  | g_B|^2 g_B    +  \frac{1}{2} M \omega_y^2  y^2   \, g_B =    \mu  \, g_B    \label{CondPsi8}\, \, . 
\end{eqnarray}
Our approach is to treat the slowly varying envelope function $\eta$ using the Thomas-Fermi approximation and impose armchair boundary conditions at the ribbon edges for the plane wave functions. To implement this method we separate the chemical potential in terms of the envelope contribution $f_e  \mu$ and the contribution from the plane-wave excitations $f_p   \mu$, where $f_e + f_p = 1$. The system Eqs.~(\ref{CondPsi7})-(\ref{CondPsi8}) splits into two pairs of equations 
\begin{eqnarray}
      \hspace{0pc}    -   \hbar c_l   g_B \frac{ \partial_y \eta}{ \eta}     +             U_\mathrm{2D} \,   \eta^2  | g_A|^2 g_A    +  \frac{1}{2} M \omega_y^2  y^2     \, g_A &=& f_e   \mu  \, g_A   \label{split1a}\, , \\
      \hspace{0pc}   \hbar c_l   g_A \frac{ \partial_y \eta}{ \eta}     +             U_\mathrm{2D} \,    \eta^2  | g_B|^2 g_B    +  \frac{1}{2} M \omega_y^2  y^2   \, g_B &=&  f_e  \mu  \, g_B    \label{split1b}\, \, , 
\end{eqnarray}
and
\begin{eqnarray}
      &&  \hbar  c_l   \left(  k_x   -  \partial_y \right)    g_B  =    f_p   \mu  \, g_A   \, , \label{split2a} \\
      && \hbar  c_l   \left(  k_x   +   \partial_y \right)    g_A   =  f_p  \mu  \, g_B    \label{split2b}\,  . 
\end{eqnarray}
Applying the Thomas-Fermi approximation in Eqs.~(\ref{split1a})-(\ref{split1b}) by neglecting derivatives of the envelope function in favor of the interaction and harmonic terms, we obtain 
\begin{eqnarray}
         \eta^2(y)  =  \frac{ f_e \mu -     (1/2) M \omega_y^2  y^2 }{  U_\mathrm{2D}    |g_A|^2 }  \, , \;\; \mathrm{and} \;\;\;  \eta^2(y)  =  \frac{ f_e \mu -     (1/2) M \omega_y^2  y^2 }{  U_\mathrm{2D}   |g_B|^2 }     \, , \label{split1c}
\end{eqnarray}
which requires the condition $|g_A| = \pm |g_B|$, for consistency. We combine Eqs.~(\ref{split2a})-(\ref{split2b}) to arrive at 
\begin{eqnarray}
       \hbar^2  c_l^2   \left(  k_x^2   -  \partial_y^2 \right)    g_B  =  f_p^2   \mu^2  \, g_B   \, , \; \;      g_A  =  \frac{1}{  f_p   \mu } \hbar  c_l   \left(  k_x   -  \partial_y \right)    g_B     \label{split2c}\,  , 
\end{eqnarray}
which we then solve by the plane wave decomposition 
\begin{eqnarray}
g_B   = b_1 e^{i k_n y} + b_2 e^{- i k_n y} \, , \label{gb}
\end{eqnarray}
with $g_A$ determined by 
\begin{eqnarray}
g_A  =  \frac{ \hbar c_l }{  f_p  \mu } \sqrt{k_x^2 + k_n^2 }  \left[   b_1\,  e^{- i \,  \alpha}   \,  e^{ i k_n y} +   b_2 \,  e^{ i \,  \alpha}   \,  e^{  - i k_n y}                     \right]  \, , \label{ga} 
\end{eqnarray}
where the phase in Eq.~(\ref{ga}) is defined as $\alpha \equiv  \mathrm{tan}^{-1} (k_n/k_x)$. We have anticipated spatially quantized states along the width of the ribbon by including the integer subscript $k_n$, $n \in \mathbb{Z}$ in Eq.~(\ref{ga}). Imposing armchair boundary conditions on $g_{A(B)}$ and their counterparts at the $\mathrm{\bf{K}}'$ point leads to the conditions 
\begin{eqnarray}
b_1 =  b_2' \;  , \;\; \; \;  b_2 = b_1' = 0 \, ,   \label{ampcond}
\end{eqnarray}
and for the wavenumber along the width of the ribbon
\begin{eqnarray}
\sin\left[ \left( k_n + K \right) L_y  \right]  = 0 \, ,\, 
\end{eqnarray}
which gives the spectral condition 
\begin{eqnarray}
k_n = \frac{n \pi }{L_y } - \frac{4 \pi}{3 a_0} \, .  \label{spectrum}
\end{eqnarray}
 In terms of the number of dimers $N$ (adjacent pairs between the ribbon edges) and the lattice spacing $a_0$, the width $L_y$ of the armchair nanoribbon is $L_y = N a_0/4$. Focusing attention on the Dirac points in the spectrum, which occur for $k_n = (4 \pi/a_0) [ (n/N) -1/3] = 0$, we find the constraint $N= 3 n$ relating the width integer $N$ and the subband quantization number $n$~\cite{Brey2006}. Thus, Dirac points occur when $N$ is an integer multiple of $3$. Combining the constraint derived from Eq.~(\ref{split1c}), namely $|g_A| = \pm |g_B|$, with Eqs.~(\ref{gb})-(\ref{ga}) and Eq.~(\ref{ampcond}) gives us the dispersion relation 
 \begin{eqnarray}
 f_p  \mu = \hbar c_l  \sqrt{k_x^2 + k_n^2} \, .  \label{dispersion2}
 \end{eqnarray}

To organize our results so far we first note that the construction of the optical lattice nanoribbon requires specifying the atomic polarizability of the atoms $\alpha$, the electric field strength of the optical lattice $E_0$, and the harmonic trap frequency $\omega_y$. For a particular choice of ribbon width $L_y$ (equivalently $N$) the envelope function Eq.~(\ref{split1c}) must vanish at the edges $\eta(\pm L_y/2) = 0$, with the edges defined along the line where the lattice potential well depth equals the magnitude of the harmonic potential: $V_\mathrm{lattice}(\pm L_y/2) =  V_\mathrm{trap}(\pm L_y/2)$. The condition for vanishing envelope implies that
\begin{eqnarray}
\eta^2(\pm L_y/2) &=&  \frac{ [ ( 1 - f_p )/f_p ] \, \hbar c_l  \, \sqrt{k_x^2 + k_n^2}   - (1/8) M \omega_y^2 L_y^2 }{U_\mathrm{2D}   |b_1|^2} = 0  \, , \label{ribcond1} 
\end{eqnarray} 
and the condition for the optical and harmonic potentials is 
\begin{eqnarray}
\frac{1}{2} M \omega_y^2 L_y^2 = \alpha E_0^2 \, .       \label{ribcond2} 
\end{eqnarray}
Combining Eqs.~(\ref{ribcond1})-(\ref{ribcond2}) determines the envelope and plane wave fractions $f_e$ and $f_p$ 
\begin{eqnarray}
f_p = 1 - f_e =  \left( \frac{\alpha E_0^2}{  4 \hbar c_l  \sqrt{k_x^2 + k_n^2} }  + 1   \right)^{-1} \, , 
\end{eqnarray}
which combines with Eq.~(\ref{dispersion2}) to give the full expression for the dispersion 
\begin{eqnarray}
   \mu = \hbar c_l  \sqrt{k_x^2 + k_n^2} + \frac{M \omega_y^2 L_y^2}{8}  \, .  \label{dispersion3}
\end{eqnarray}
The first term in Eq.~(\ref{dispersion3}) arrises from excitations in the long and short directions of the ribbon and the second term accounts for the finite width of the ribbon, which in practicality may be subtracted off as an overall constant energy by defining the shifted chemical potential $\mu  \to  \mu - M \omega_y^2 L_y^2/8$. Upon inclusion of the $\mathrm{\bf{K}}'$-point contribution, the 2-component spinor wavefunction is given by 
\begin{eqnarray}
\hspace{-6pc}  \Psi(x, y, t) =    \\
 \hspace{-6pc}   e^{i \left( k_x x - \mu \, t/\hbar \right) }  \,     \sqrt{ \frac{M   \omega_y^2  L_y^2 }{ 2 U_\mathrm{2D}     }  \left( 1 - 4 \frac{y^2}{L_y^2}  \right) }   \left(  \! \begin{array}{c} 
                      \mathrm{cos}\left(k_n \, y \right)      \\
                  \pm   \frac{  i }{ \sqrt{k_x^2 + k_n^2}}       \left[ k_x  \mathrm{cos}\left(k_n \, y \right)     +  k_n  \mathrm{sin}\left(k_n \, y \right)     \right]                \end{array} \!  \right) \, ,                
\end{eqnarray}
defined over the width $-L_y/2 < y< + L_y/2$. In particular, for the linear degenerate subband where $k_n =0$ the wavefunction reduces to
\begin{eqnarray}
\hspace{-2pc}  \Psi(x, y, t) =  e^{i \left( k_x x - \mu \, t/\hbar \right) }  \,     \sqrt{ \frac{M   \omega_y^2  L_y^2 }{ 2 U_\mathrm{2D}     }  \left( 1 - 4 \frac{y^2}{L_y^2}  \right) }   \left( \begin{array}{c} 
                      1      \\
                     \pm  i    \end{array} \right) \, .     \label{factorizedpsi}
\end{eqnarray}

In the remainder of this section we use our results up to this point to derive the quasi-1D NLDE, Eqs.~(\ref{eqn:CondPsi7})-(\ref{eqn:CondPsi8}) and Eqs.~(\ref{eqn:CondPsi9})-(\ref{eqn:CondPsi10}), starting from the 2D NLDE in Eqs.~(\ref{eqn:CondPsi5})-(\ref{eqn:CondPsi6}). Transforming to the quasi-one-dimensional regime requires that $L_y \ll L_x$, which ensures that excitations along the $x$-direction have much lower energy than those in the $y$-direction. In particular, this condition must be satisfied for the armchair nanoribbon geometry discussed thus far. Equivalently, these constraints may be expressed in terms of the trap frequencies and the atom-atom interaction: $\omega_x \ll \omega_y$, $\hbar \omega_x \ll U$. For the moment we neglect the harmonic trap in $x$.  A modified renormalized interaction $U$ and careful consideration of phase coherence in quasi-1D BECs should be sufficient to account for tight confinement, but are not necessary here~\cite{Olshanii1998,Petrov2000,Petrov2001,Cristiani2002}.

The quasi-1D NLDE is obtained by separating $\psi_A({\bf r},t )$ and $\psi_B({\bf r},t )$ into longitudinal and transverse modes following similar arguments as in Ref.~\cite{Carr2000}: 
\begin{eqnarray}
\psi_A({\bf r},t )  =  \left( \frac{M \omega_y^2 L_y^3}{3 U_{2D} } \right)^{-1/2}     h(y) f_A(x) e^{-i \mu t/\hbar} \, ,\label{split1} \\
 \psi_B({\bf r}, t)  =   \left( \frac{M \omega_y^2 L_y^3}{3 U_{2D} } \right)^{-1/2}    h(y) f_B(x)  e^{-i \mu t/\hbar} \, , \label{split2}
\end{eqnarray}
where $f_{A(B)}(x)$ contains the longitudinal $x$-dependence and $h(y)$ is the dimensionless function that describes the transverse part of the wavefunction from Eq.~(\ref{factorizedpsi}), i.e., 
\begin{eqnarray}
h(y) =   \sqrt{ \frac{M   \omega_y^2  L_y^2 }{ 2 U_\mathrm{2D}     }  \left( 1 - 4 \frac{y^2}{L_y^2}  \right) }   \, . \label{hfunction}
\end{eqnarray}
Note that we have included a normalization prefactor in Eqs.~(\ref{split1})-(\ref{split2}). Substituting Eqs.~(\ref{split1})-(\ref{split2}) into Eqs.~(\ref{eqn:CondPsi5})-(\ref{eqn:CondPsi6}) using the expression for $h(y)$ in Eq.~(\ref{hfunction}) and integrating over the $y$-direction leads directly to the quasi-1D NLDE with complex coefficients in Eqs.~(\ref{eqn:CondPsi7})-(\ref{eqn:CondPsi8}). Consequently, we find the quasi-1D renormalized interaction to be 
\begin{eqnarray}
 U_\mathrm{1D} &\equiv  U_\mathrm{2D}   \left( \frac{3  }{2 L_y }\right)^{2} \hspace{-.5pc}   \int_{- L_y/2}^{+  L_y/2 } \hspace{-.5pc} \! dy \, \left(   1  - 4  \frac{y^2}{L_y^2 } \right)   \\
 &=   \left( \frac{6 }{ 5 L_y }\right)  U_\mathrm{2D}   \, .
\end{eqnarray}

A key result here is that the chemical potential is not modified by dimensional reduction since the Dirac equation is first order in the spatial derivatives and the extra term proportional to $dh/dy$ is an antisymmetric function of $y$ which vanishes upon integration. To simplify the notation, for the rest of this article we will use the plain notation $U$ and understand that this refers to the quasi-one-dimensional renormalized interaction.

\section{General properties of NLDE solutions: fixed points and invariance relations}
\label{SolutionProperties}

As a first step towards solving the NLDE, we map out the solution landscape by understanding the character of the various solution types. In this section and throughout our work we confine our analysis to stationary solutions, leaving the case of dynamics for future investigations. In particular, we require a clear understanding of the points where solutions are constant (zero spatial derivative), and the flow of solutions near these fixed points. Working from Eqs.~(\ref{eqn:CondPsi9})-(\ref{eqn:CondPsi10}) we look for real solutions and write the NLDE as a derivative field (or direction field)
\begin{eqnarray}
   f_B'   =  -\frac{U}{\hbar c_l} f_A \left(  \frac{\mu}{U}  -  f_A^2 \right)  ,  \label{zigzag1} \\
   f_A'  =   \frac{U}{\hbar c_l} f_B   \left(  \frac{\mu}{U}  -  f_B^2 \right)  \label{zigzag2}  \, , 
\end{eqnarray}
where the dependence on $x$ is implied. Together, Eqs.~(\ref{zigzag1})-(\ref{zigzag2}) comprise a vector field $(f_A', \, f_B')$ which describes the flow of two-spinor solutions. The various combinations of conditions on the signs of $f_A'$ and $f_B'$ partition the $(f_A, \, f_B)$ solution space into 16 regions. However, analysis of symmetries of Eqs.~(\ref{zigzag1})-(\ref{zigzag2}) shows that only 8 combinations lead to distinct solution types. In particular, the transformation $f_A \to - f_B, \; f_B \to f_A$ leaves our equations invariant. We have listed these regions in Table~\ref{table1} along with the corresponding signs for the derivatives $f_A^\prime$ and $f_B^\prime$. Nine fixed points exist in the solution space $(f_A, \, f_B)$: $(0, 0), \, (\pm \sqrt{\mu/U}, \, 0), \, (\pm \sqrt{\mu/U}, \, \pm \sqrt{\mu/U} ), \, ( 0 , \, \pm \sqrt{\mu/U} )$.

\begin{table}[h]
 \begin{indented}
\item[]\begin{tabular}{@{\hspace{.75pc}}llllll }
\br 
\lineup
 Region &  Condition for  $f_A$  &  Condition for  $f_B$   & $f_A^\prime$   & $f_B^\prime$   & Solution type  \\
 \mr
 I &    $  \sqrt{\mu/U} < f_A $  &  $\sqrt{\mu/U} <  f_B$& $ -$ & $+$ &   bright, multi-soliton        \\

 II   &  $  \sqrt{\mu/U} <  f_A$ &  $ 0<   f_B< \sqrt{\mu/U}$ & $+$  & $+$ &   bright, multi-soliton      \\

  III &  $  \sqrt{\mu/U} <  f_A $  & $ - \sqrt{\mu/U} < f_B < 0$  &  $-$  & $+$  &   multi-soliton      \\
 
   IV &  $  \sqrt{\mu/U} <  f_A$  &  $ f_B<  - \sqrt{\mu/U}$ & $+$  & $+$   &  multi-soliton            \\
 
    V & $ 0<  f_A  < \sqrt{\mu/U}$             & $  \sqrt{\mu/U}<  f_B  $   & $-$  & $-$  &   bright, multi-soliton        \\
   
     VI & $ 0<  f_A < \sqrt{\mu/U}$  & $  0  < f_B <  \sqrt{\mu/U}$ & $+$   & $-$   &   dark soliton        \\
      VII & $ 0<  f_A < \sqrt{\mu/U}$  &    $ - \sqrt{\mu/U} < f_B < 0$  & $-$   & $-$ &           \\
       VIII & $ 0<  f_A < \sqrt{\mu/U}$  & $ f_B<  - \sqrt{\mu/U}$ & $+$   & $-$ &    multi-soliton              \\
 \br 
\end{tabular}   
{\caption{\emph{Distinct solution regions of the NLDE.} The sign of each derivative is determined by the values of both spinor functions. Note that in each region the signs of $f_A^\prime$ and  $f_B^\prime$ remain fixed, where a sign change occurs across a boundary. The boundary between two regions is defined by $f_A^\prime =0$ or $f_B^\prime =0$, which corresponds to replacing an inequality by equality in any of the conditions listed above. The far right column lists the three types of soliton solutions and their associated regions.}   \label{table1}}
 \end{indented}
\end{table}

A qualitative analysis of each region in Table~\ref{table1} provides a guide to the types of soliton. Two main types of solutions exist: oscillating solutions which do not flatten out asymptotically, and localized solutions whose derivatives vanish asymptotically. The former turn out to be variants on multi-solitons, while the latter turn out to be varieties of single solitons. In addition, solutions differ qualitatively depending on which regions in Table~\ref{table1} are involved.

\begin{figure}[h]
\centering
\subfigure{
\label{fig:ex3-a}
\includegraphics[width=.5\textwidth]{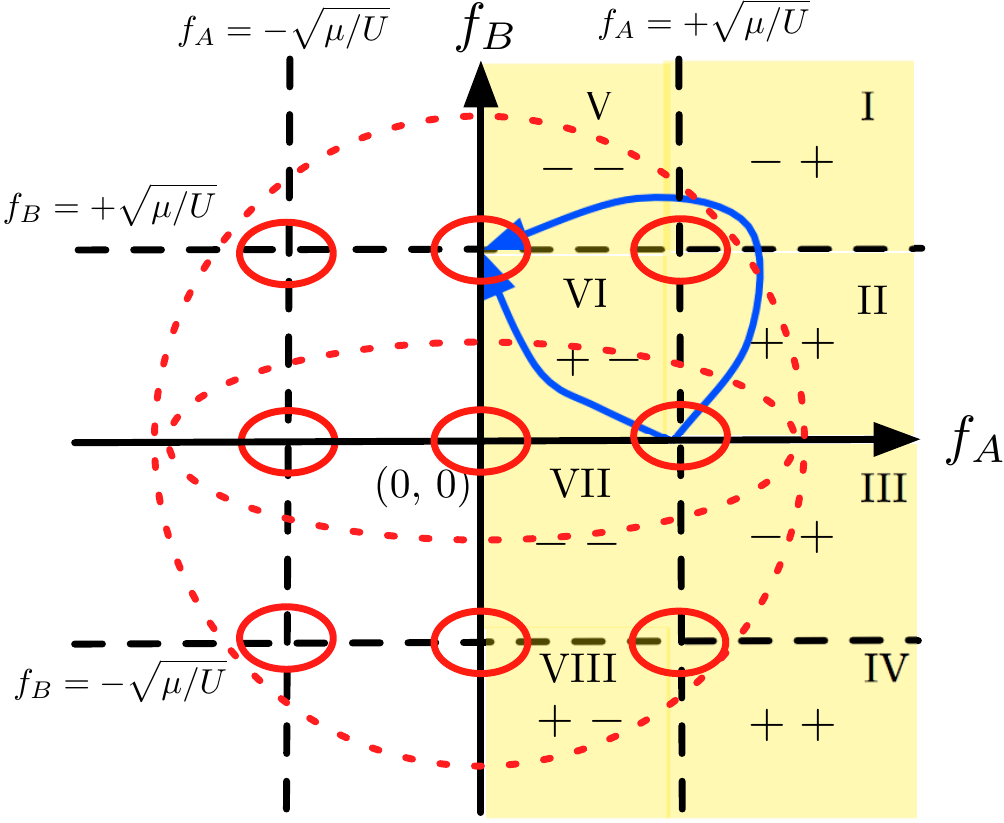}}\\ 
\caption[]{\emph{Character of NLDE solutions and regions}. Small amplitude oscillating solutions (solid red) are shown centered on fixed points, large amplitude oscillations encircling several fixed points (dashed red), and asymptotically flat solitons (blue) interpolating between fixed points. Regions of negative $f_A$ can be obtained by a trivial transform as stated in text. The regions described in Table~\ref{table1} are highlighted in yellow.} \label{SolutionTypes}
\end{figure}

At one extreme, strongly oscillating (periodic) solutions exist for which $f_A$ and $f_B$ both have amplitudes greater than $\sqrt{\mu/U}$, in which case each component has three critical points ($f_{A(B)}^\prime=0$) during a half period. Such solutions cover all regions in Table~\ref{table1} except regions VI and VII. We give an example of this type of solution in this section, which we obtain analytically. Other oscillating solutions occur around each of the fixed points $( \pm \sqrt{\mu/U}, \, 0)$,  which may be discerned from Eqs.~(\ref{zigzag1})-(\ref{zigzag2}) using linear perturbation theory by substituting $f_A(x) =  \epsilon \, \mathrm{cos} k x \pm \sqrt{\mu/U}$ or  $f_A(x) = \epsilon \, \mathrm{cos} k x$, with the small amplitude $\epsilon$ such that $\epsilon/ \sqrt{\mu/U} \ll 1$. These solutions cross four region boundaries. For example, regions II, III, VI, and VII in the case where $f_A(x) =  \epsilon \, \mathrm{cos} k x \pm \sqrt{\mu/U}$, with each derivative $f_A'$ and $f_B'$ changing sign once per half period. We also find non-oscillating asymptotically flat solutions having a similar form to the function $(1/2)[1\pm \mathrm{tanh}(x)]$, which we study in Sec.~\ref{GeneralSolutions}. This spinor solution has a constant total density everywhere except near a localized region where a dip, or notch, in the density occurs, a form which describes a dark soliton. In addition, we find that a bright soliton which crosses regions I, II, and V.  Figure~\ref{SolutionTypes} gives a qualitative schematic depiction of the various regions, oscillating solutions centered on fixed points, and asymptotically flat solutions which interpolate between fixed points.

To deepen our analysis, a detailed map of the solution space of Eqs.~(\ref{zigzag1})-(\ref{zigzag2}) can be arrived at by uncovering spatially invariant quantities. To obtain the first invariant quantity we multiply Eq.~(\ref{zigzag1}) by the right hand side of Eq.~(\ref{zigzag2}) and vice versa, then add the resulting equations to obtain 
\begin{eqnarray}
  \left(  \mu f_A  - U \, f_A^3 \right)   f_A^\prime +   \left(  \mu f_B -  U  f_B^3 \right)  f_B^\prime  = 0   \label{zigzag3}  \, , 
\end{eqnarray}
which then gives
\begin{eqnarray}
 \left[ \left( 2 \mu  f_A^2   - U   f_A^4 \right)    +   \left( 2  \mu  f_B^2   - U   f_B^4   \right)  \right]^\prime  = 0   \label{zigzag3}  \, , 
\end{eqnarray}
where the prime notation indicates differentiation with respect to $x$. Integrating Eq.~(\ref{zigzag3}) gives a relation between the functions $f_A$ and $f_B$
\begin{eqnarray}
\left( f_A^2  - \frac{\mu}{U} \right)^2 + \left( f_B^2  - \frac{\mu}{U} \right)^2 - 2\left(  \frac{\mu}{U} \right)^2 = C \, ,\label{firstconservation}
\end{eqnarray}
where we have simplified the expression by completing the squares in $f_A$ and $f_B$, and $C$ is the integration constant. The meaning of Eq.~(\ref{zigzag3}) is seen by multiplying Eq.~(\ref{zigzag1}) by $f_B$ and Eq.~(\ref{zigzag2}) by $f_A$ and then adding the resulting equations, which gives 
\begin{eqnarray}
\hspace{-2pc} -  \frac{U}{2}  \left[    \left(  f _A^4 +  f_B^4 \right)  -  2 \frac{\mu}{U} \left(   f_A^2+ f_B^2 \right)   \right] = \hbar c_l \left( f_A^\prime  f_B - f_A f_B^\prime  \right)  +  \frac{U}{2}  \left(  f _A^4 +  f_B^4 \right)   \, . \label{nldesum}
\end{eqnarray}
The expression in brackets on the left of Eq.~(\ref{nldesum}) is the same as the left hand side of Eq.~(\ref{firstconservation}), while the right side of Eq.~(\ref{nldesum}) is the $T^{11}$ element of the energy-momentum tensor $T^{\mu \nu}$~\cite{Saxena2012}. Thus, we find 
\begin{eqnarray}
T^{11} = - \frac{U C}{2} \, , \label{Tdef}
\end{eqnarray}
so that Eq.~(\ref{firstconservation}) is a statement of uniformity of pressure $T^{11}/c_l$ in the $x$-direction. The other elements of the energy-momentum tensor can also be computed, whereby we obtain the total energy density $T^{00} = -(\hbar c_l/2)  (f_A^\prime f_B - f_A f_B^\prime ) +  (U/2) (f_A^4 + f_B^4) $ and current $T^{01} = T^{10} = (i/2) ( \Psi^\dagger \Psi^\prime -  {\Psi^\dagger}^\prime \Psi   ) = 0$. Note that the current vanishes but the energy density is spatially dependent. Furthermore, the energy density can be expressed in terms of the pressure $T^{11}$ and the interaction energy as $T^{00} = - (1/2) T^{11} + (3/4) U(f_A^4 + f_B^4)$.

A second quantity which characterizes solutions of the NLDE is arrived at by expressing Eqs.~(\ref{zigzag1})-(\ref{zigzag2}) in the form 
\begin{eqnarray}
  f_B  f_B^\prime  =  -\frac{U}{\hbar c_l}f_B f_A \left(  \frac{\mu}{U}  -  f_A^2 \right)  ,  \label{fBfA} \\
   f_A f_A^\prime  =   \frac{U}{\hbar c_l} f_A f_B  \left(  \frac{\mu}{U}  -  f_B^2 \right)  \label{fAfB}  \, , 
\end{eqnarray}
then adding, integrating, and combining terms to get a total derivative on the left hand side:
\begin{eqnarray}
 \left( f_A^2+  f_B^2 \right)^\prime =  \frac{2 U}{\hbar c_l} f_Bf_A\left(    f_A^2  - f_B^2 \right) .\label{totaldensity} 
\end{eqnarray}
Equation~(\ref{totaldensity}) can be simplified by introducing the average spin components $S_{x} \equiv  \bar{\Psi} \sigma_{x} \Psi = 2 f_A f_B $, $S_{z} \equiv  \bar{\Psi} \sigma_{z} \Psi = 2 f_A^2 -  f_B^2$, and the total density $\rho  \equiv f_A^2  + f_B^2$, so that Eq.~(\ref{totaldensity}) becomes $\rho^\prime =  \left( U/\hbar c_l \right)  S_x S_z$. Equation~(\ref{totaldensity}) states that the total density varies most where the wavefunction lies between a pure chiral state and a highly mixed state, as $S_z$ is a measure of chirality and $S_x$ a measure of the degree of chiral mixing within a particular state. Integrating Eq.~(\ref{totaldensity}) over an interval $a < x < b$ gives
\begin{eqnarray}
     \int_a^b \! dx\,  S_x S_z =  \frac{\hbar c_l}{U} \left[  \rho(b) - \rho(a) \right] \,.   \label{conservedcharge}
\end{eqnarray}
In particular, Eq.~(\ref{conservedcharge}) allows for solutions which asymptotically approach a constant value, i.e., 
\begin{eqnarray}
    \int_{- \infty}^{+\infty}   \! dx\,  S_x S_z   = \delta_{\pm \infty}\, , \label{conservedcharge2}
\end{eqnarray}
where the constant $\delta_{\pm \infty} \equiv (\hbar c_l/U) \left[ \rho(+\infty ) - \rho(-\infty)\right]$ is a global parameter that depends only on the difference between the asymptotic values of the total density. Hence, stationary solitons are topologically stable in that local fluctuations in $f_A$ and $f_B$ will cancel out, with $\delta_{\pm \infty}$ remaining fixed.

\section{Transition of solutions across the soliton boundary}
\label{GeneralSolutions}

We now combine the qualitative information and the invariance relation (Eq.~(\ref{firstconservation})) from Sec.~\ref{SolutionProperties} to arrive at a more technical understanding  of the relationship between oscillating and asymptotically flat solutions of the NLDE. In particular, by choosing specific values for the chemical potential $\mu$ and strength of nonlinearity $U$, Eqs.~(\ref{zigzag1})-(\ref{zigzag2}) may be solved by numerical or analytical methods. In general though, such solutions may be highly oscillatory and not necessarily stable. In this section we study the evolution of oscillating solutions as $\mu$ and $U$ are tuned to obtain particular soliton solutions which are stable. 
\begin{figure}[h]
\centering
\subfigure{
\label{fig:ex3-a}
\includegraphics[width=.65\textwidth]{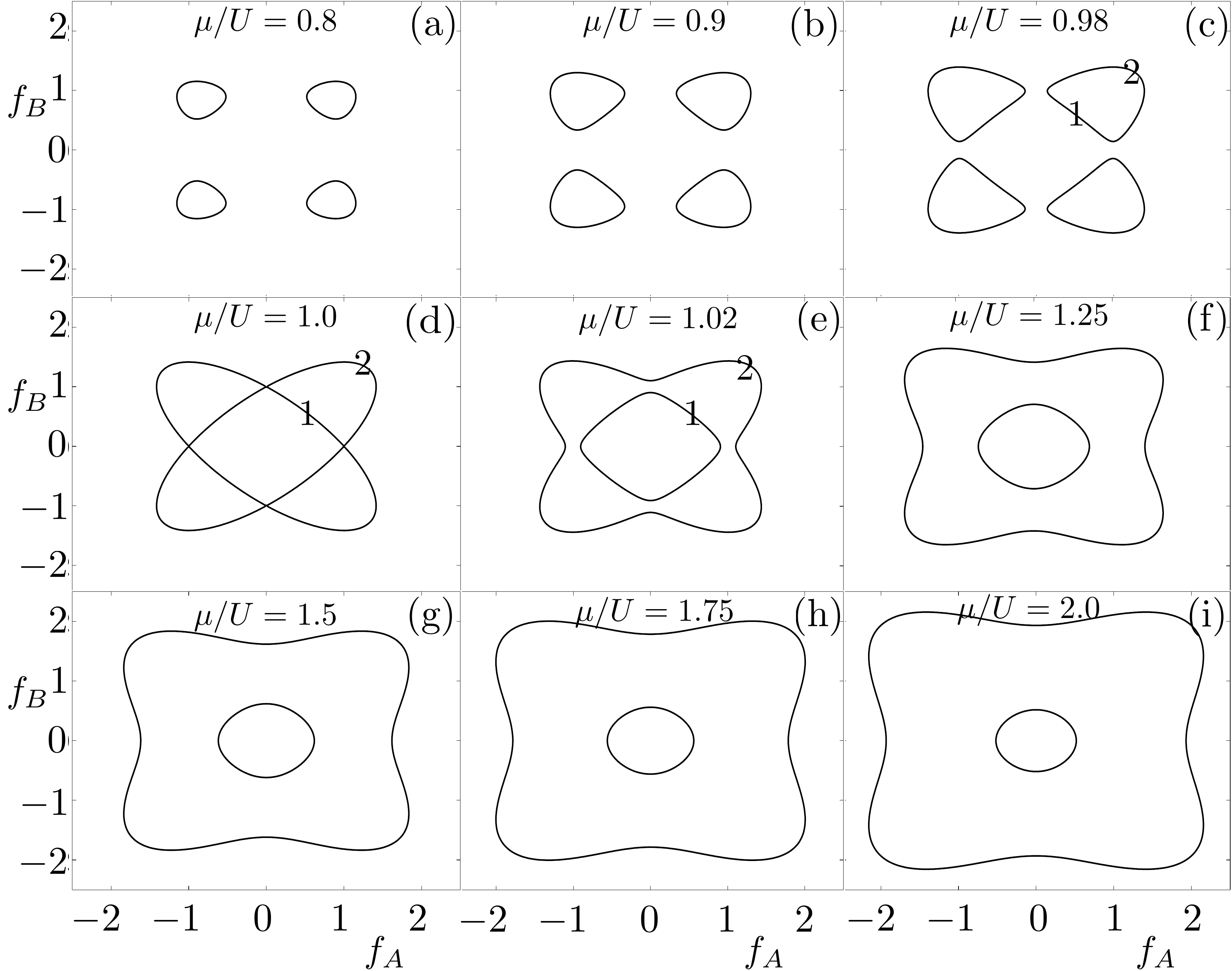}} \\
\caption[]{\emph{Evolution of NLDE solution orbits in parameter space}. For a chosen value of the fixed quantity, $T^{11}= \mu^2/2U$, orbits evolve as the ratio of chemical potential to interaction, $\mu/U$, is tuned from values less than one to greater than one. For the particular value $\mu/U =1$, shown in (d), solutions labeled as paths 1 and 2 begin and end on the fixed points $(0,1)$ and $(1, 0)$ (see Table~\ref{table1}). Solutions must approach a fixed point asymptotically for $x \to + \infty, \, -\infty$, so that the non-trivial (nonzero spatial derivative) part of the solution is spatially localized. Paths 1 and 2 describe two distinct solitons and divide the solution space into qualitatively distinct regions: in (a)-(c), the solution oscillates about one of four fixed points $(\pm 1, \pm 1)$; in (e)-(i), a large amplitude solution encircles all nine fixed points, and a small amplitude solution oscillates about the fixed point at $(0, 0)$.}
\label{orbits1}
\end{figure}

We may better understand solutions of Eqs.~(\ref{zigzag1})-(\ref{zigzag2}) by plotting Eq.~(\ref{firstconservation}) for a particular value of $T^{11}$. This allows us to see how solutions evolve as we vary the ratio $\mu/U$. For the choice $T^{11}= \mu^2/2U$, which corresponds to $C= -(\mu/U)^2$ in Eq.~(\ref{firstconservation}), we have plotted the solution space for $\mu/U = 0.8$, $0.9$, $0.98$, $1.0$, $1.02$, $1.25$, $1.5$, $1.75$, $2.0$, in Figs.~\ref{orbits1}(a)-(i). In Fig.~\ref{orbits1}(a)-(c), solutions oscillate about the fixed points $(f_A, f_B) = (\pm \sqrt{\mu/U}, \pm \sqrt{\mu/U})$ with the orbits beginning to coalesce in Fig.~\ref{orbits1}(c). In Fig.~\ref{orbits1}(d), $\mu/U=1$ and solution paths begin and end at the saddle fixed points $(0, \pm \sqrt{\mu/U}),\, (\pm \sqrt{\mu/U}, 0 )$, asymptotically flattening out for large positive and negative $x$. In this case there are two distinct types of solitons indicated by the paths $1$ and $2$ that connect the points $(1,0)$ and $(0,1)$. We shall see that these solutions correspond to dark and bright solitons as previously mentioned. As $\mu/U$ is increased from unity, orbits bifurcate into solutions which oscillate about $(0,0)$, with small and large amplitudes indicated by the paths $1$ and $2$ in Fig.~\ref{orbits1}(e). These orbits continue to smooth out through Fig.~\ref{orbits1}(i), at which point $\mu/U = 2$. Soliton solutions such as those depicted in Fig.~\ref{orbits1}(d) correspond to the case $T^{11}= \mu^2/2U$. In general though, isolated dark and bright single or multi-solitons are distinguished by the density conditions 
\begin{eqnarray}
  0  <  \rho_\mathrm{DS} <    \frac{ \mu}{U} \,\bar{n}  \; , \hspace{2pc}   \frac{\mu}{U} \, \bar{n}    <  \rho_\mathrm{BS}   <     \frac{\mu}{U} \left[ 1 + \sqrt{2 \left( 1 - \frac{T^{11} U}{\mu^2} \right)  } \right]   \bar{n} .  \label{densityconditions}
\end{eqnarray}
The upper and lower bounds $\mu \bar{n}/U$ result from Eqs.~(\ref{zigzag1})-(\ref{zigzag2}), and the upper bound on the second inequality comes from Eqs.~(\ref{firstconservation}) and (\ref{Tdef}). We have added the subscripts DS and BS to the total density in Eq.~(\ref{densityconditions}) to indicate dark and bright solitons, respectively. Note the inclusion of the average particle density $\bar{n}$. The densities are defined in terms of the real spinor spatial functions $\rho_\mathrm{DS}(x) = \bar{n} \left[  f^2_{A, \mathrm{DS}}(x)  + f^2_{B, \mathrm{DS}}(x)\right]$, with an analogous definition for the bright soliton.

The bounds for the inequalities in Eq.~(\ref{densityconditions}) can be proved as follows. For the proof of the upper bound of the first inequality, we start by assuming that $\rho_\mathrm{DS}(x_1) \ge  \bar{n}\mu /U$ for at least one element in the domain $x_1 \in \mathbb{R}$, and where $ \bar{n}\mu /U > 0$. Since we are considering isolated dark solitons, there exist an infinite number of points $x_2 \in \mathbb{R}$ such that $\rho_\mathrm{DS}(x_1) < \rho_\mathrm{DS}(x_2)$. Moreover, we have $\lim_{\, x \to \pm \infty}   \rho_\mathrm{DS}(x) =   \rho_\mathrm{DS}^{\mathrm{sup }} \equiv  \mathrm{sup}\left\{ \rho_\mathrm{DS}(x)  \in \mathbb{R}: x \in \mathbb{R}) \right\} $, where $\rho_\mathrm{DS}^{\mathrm{sup }}$ denotes the supremum of the dark soliton density. Thus, $\bar{n} \mu/U <  \rho_\mathrm{DS}^{\mathrm{sup}}$ and $\lim_{\, x \to \pm \infty}   \rho_\mathrm{DS}^\prime(x) = 0$. It follows that $\lim_{\, x \to \pm \infty} \left( f_{A, \mathrm{DS}}^2 + f_{B,\mathrm{DS}}^2 \right)  =  \rho_\mathrm{DS}^{\mathrm{sup }}/\bar{n}$ $\Rightarrow$ $- \sqrt{  \rho_\mathrm{DS}^{\mathrm{sup }}/\bar{n}} <    f_{A(B), \mathrm{DS}} <  + \sqrt{  \rho_\mathrm{DS}^{\mathrm{sup }}/\bar{n}}$, and that $\lim_{\, x \to \pm \infty}  f^\prime_{A(B), \mathrm{DS}}(x) = 0$. Here the subscript indicates that the condition applies to both $f_A$ and $f_B$. By Eqs.~(\ref{zigzag1})-(\ref{zigzag2}), asymptotically vanishing derivatives imply four possible combinations: $\lim_{\, x \to \pm \infty}  \left\{  f_{A, \mathrm{DS}}^2(x),    f_{B, \mathrm{DS}}^2(x)      \right\} = \left\{ 0,0\right\},  \left\{ \mu/U ,0\right\} , \left\{ 0, \mu/U \right\} ,  \left\{ \mu/U , \mu/U   \right\}$. The first three cases lead to $\rho_\mathrm{DS}^{\mathrm{sup } }= 0, \bar{n} \mu/U,  \bar{n} \mu/U$, respectively, which contradict our earlier result that $    0 < \bar{n} \mu/U   < \rho_\mathrm{DS}^{\mathrm{sup } }$. The fourth combination cannot occur as one may deduce from Eqs.~(\ref{zigzag1})-(\ref{zigzag2}) and the analysis in Table~\ref{table1}, proving by contradiction the upper bound of the first inequality in Eq.~(\ref{densityconditions}). Proving the lower bound in the second inequality proceeds by the reverse argument, i.e., using the infimum $\rho_\mathrm{BS}^{\mathrm{inf}}$ and the initial assumption that $\rho_\mathrm{DS}(x_1) \le  \bar{n}\mu /U$, but otherwise the steps are similar to those in the first proof. Finally we address the upper bound in the second inequality in Eq.~(\ref{densityconditions}). This bound comes from the invariance relation Eq.~(\ref{firstconservation}) where one sees that the sum of squared terms places an upper bound on $f_{A,\mathrm{BS}}(x)$ and $f_{B,\mathrm{BS}}(x)$, and thus on the total density $\rho_\mathrm{BS}(x)$. We obtain this result by setting $f_{A,\mathrm{BS}}(x)$ or $f_{B,\mathrm{BS}}(x)$ equal to zero in Eq.~(\ref{firstconservation}) and using Eq.~(\ref{Tdef}) to replace $C$ by $-T^{11}U/2$. Note that the radical in Eq.~(\ref{densityconditions}) implies the upper bound $T^{11}_\mathrm{max} = \mu^2/U$.

Thus, solitons exist on a 3-dimensional sub-manifold of parameters (defined by the condition $\mu =  |2T^{11} U|^{1/2} $) of the 4-dimensional parameter manifold determined by the chemical potential, interaction, density, and energy-momentum tensor with coordinates denoted as $(\mu , \, U, \rho, \, T^{11})$. Moreover, the density conditions in Eq.~(\ref{densityconditions}) further partition the 3D parameter subspace into the two types of solitons along the boundary $\rho = \bar{n} |2T^{11}/U|^{1/2} = \bar{n}  \mu/U$, where the last equality pertains to the spacial case $T^{11} = \mu^2/2U$.  

\begin{figure}[t]
\centering
\subfigure{
\label{fig:ex3-a}
 \includegraphics[width=.75\textwidth]{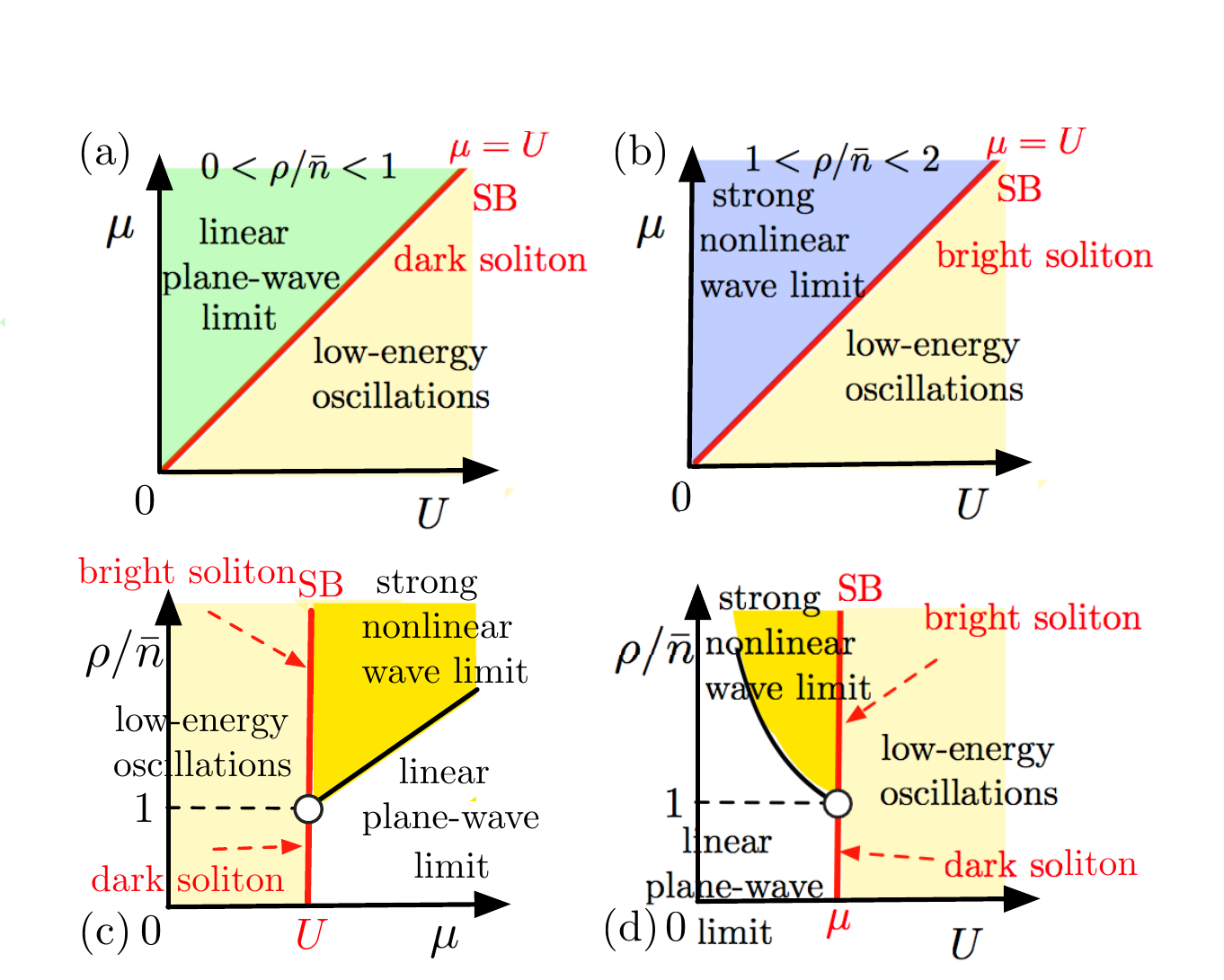}} \\
\caption[]{\emph{Quantum phase transitions across the soliton boundary}. The soliton boundary (SB) is depicted as the bold red line in each figure. (a) Solution types for low densities in the chemical potential versus interaction plane, and (b) for large densities. The soliton boundary coincides with the line $\mu =U$ in both cases. For increasing $\mu$ or decreasing $U$ across the boundary, solutions bifurcate from relatively low energy oscillations into linear plane waves or strong nonlinear waves depending on the value of the local density. Bifurcations in mean field theory indicate quantum phase transitions in the underlying many body physics (see Ref.~\cite{Ueda2009}). (c) Density versus chemical potential diagram. (d) Density versus interaction diagram with $\rho \propto  1/U$ on the boundary between the nonlinear and linear wave limit. Note that the open points $(U, 1)$ and $(\mu, 1)$ correspond to the spatially trivial solution $f_A = f_B = 1$.}
\label{phasetransition}
\end{figure}

The concept of a soliton boundary is useful in order to visualize the transition from oscillating solutions at weak nonlinearity into oscillating solutions at strong nonlinearity, with single isolated solitons appearing at the boundary between the two oscillating regimes. In particular, for $ T^{11}= \mu/2$ the soliton boundary occurs at $(\mu/U)_\mathrm{SB} = 1$. Tuning $\mu/U \to 1$ while keeping $\rho < \bar{n}$ forces the solution to the dark soliton, whereas tuning $\mu/U \to1$ while keeping $\bar{n} < \rho < 2 \bar{n}$ converges on the bright soliton. It is important to keep in mind that the upper bound here, $\rho < 2 \bar{n}$, comes from choosing a particular value for $T^{11}$ and that bright solitons exist at higher densities but are associated with a different choice of $T^{11}$. Note that the type of soliton obtained is independent of whether $\mu/U$ approaches $1$ from above or below. Conversely, if we maintain the condition $\mu/U =1$ while tuning $\rho$ through the critical value $\rho_c = \bar{n}$ we induce a transition between the dark soliton and bright soliton. Thus, tuning $\mu/U$ moves the system between oscillating regimes across the soliton boundary, while tuning $\rho/\bar{n}$ along $(\mu/U)_\mathrm{SB}$ moves the system between the dark and bright solitons. Quantum phase transitions across the soliton boundary are summarized in Fig.~\ref{phasetransition}. This kind of phase transition in the mean-field theory indicates a possible corresponding quantum phase transition in the underlying microscopic theory~\cite{Ueda2008,Ueda2009,Ueda2010,Kanamoto2010}. A more exact phase diagram could be calculated from the many body theory via the RLSE, forming a subject for future work. For contemporary works on quantum phase transitions see Ref.~\cite{CarrBook2010}.

\begin{figure}[t]
\centering
\subfigure{
\label{fig:ex3-a}
\includegraphics[width=.6\textwidth]{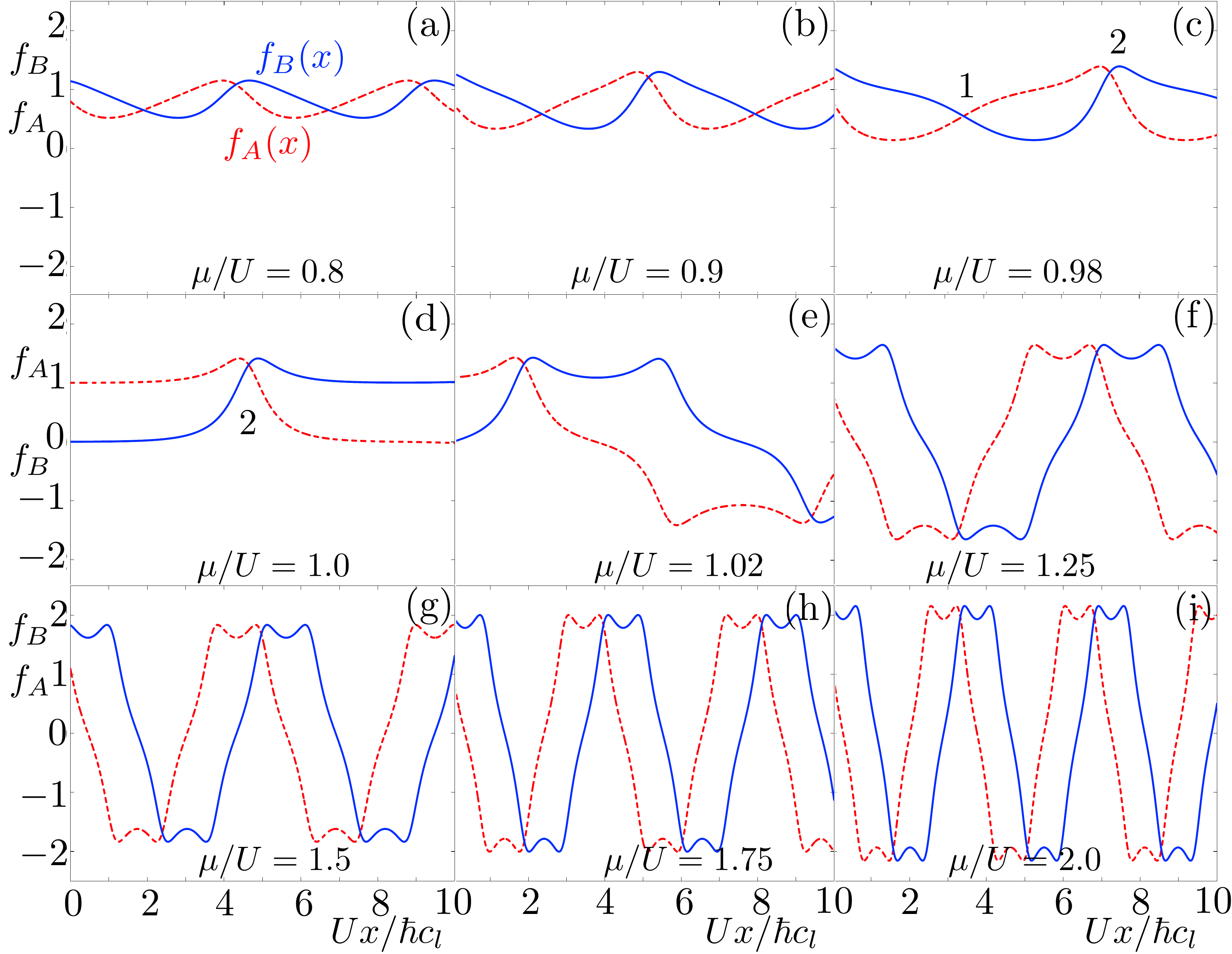}} \\
\caption[]{\emph{Convergence to a single bright soliton}. (a)-(c) Transition into the soliton boundary focusing on the bright soliton shown in (d). (e)-(i) Transition away from the soliton boundary.}
\label{orbits2}
\end{figure}

\begin{figure}[t]
\centering
\subfigure{
\label{fig:ex3-a}
\includegraphics[width=.6\textwidth]{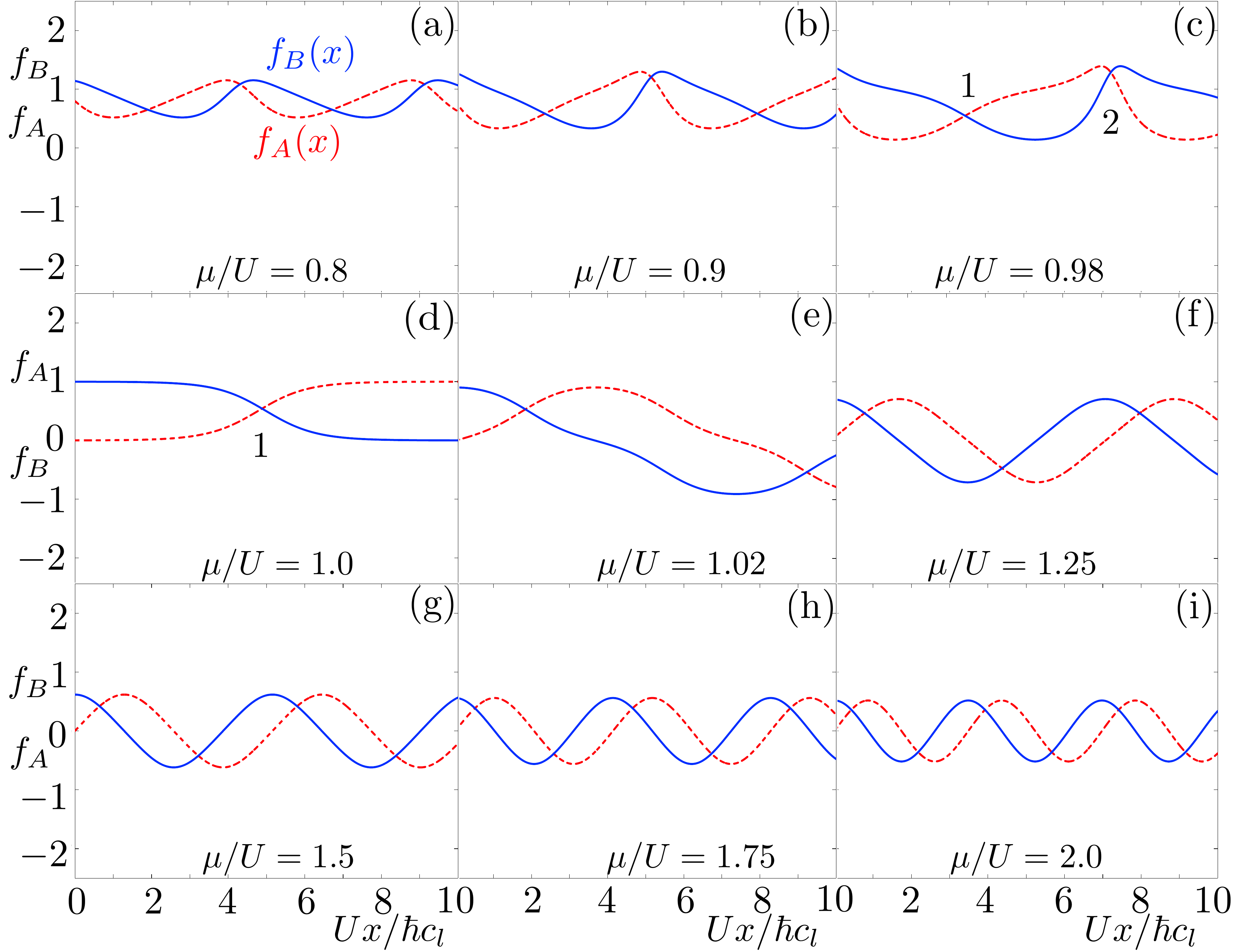}} \\
\caption[]{\emph{Convergence to a single dark soliton}. (a)-(c) Transition into the soliton boundary focusing on the dark soliton shown in (d). (e)-(i) Transition away from the soliton boundary.}
\label{orbits3}
\end{figure}

To make our analysis more concrete, we solve the numerical initial value problem defined by Eqs.~(\ref{zigzag1})-(\ref{zigzag2}) with the initial conditions taken from Eq.~(\ref{firstconservation}), which relates $f_A$ and $f_B$ at $x=0$. The value of $f_A(0)$ is chosen to coincide with a particular branch and a three-point balanced finite difference scheme for the first derivative is implemented. In Figs.~\ref{orbits2}(a)-(i), we have plotted the evolution of solutions corresponding to the upper branch (labeled as branch 2) of the orbits in Fig.~\ref{orbits1} using the initial values $(f_A(0), f_B(0)) = (0.8000,1.1421)$, $(0.8000,1.2819)$, $(0.8000,1.3702)$, $(1.0010,0.0014)$, $(1.0100,0.0000)$, $(1.1000, 1.6454)$, $(1.1000, 1.8298)$, $(1.1000, 1.9871)$, $(1.1000, 2.1272)$. The values of $f_A(0)$ were chosen to pick out branch 2 with $f_B(0)$ determined by inverting and solving Eq.~(\ref{firstconservation}) for $f_B$
\begin{eqnarray}
 f_B(0) = \pm \left[  \mu/U \pm \sqrt{ 2T^{11}/U+ 2 (\mu/U)^2 - \left( f_A(0)^2 - \mu/U \right)^2} \right]^{1/2}   .   \label{fBinitialcond} 
\end{eqnarray}
Similar plots focusing on branch 1 are shown in Fig.~\ref{orbits3} using the initial values: $(f_A(0), f_B(0)) = (0.8000, 1.1421)$, $(0.8000, 1.2819)$, $(0.8000, 1.3702)$, $(0.001, 0.9993)$, $(0.001, 0.9050)$, $(0.001, 0.7071)$, $(0.001, 0.6180)$, $(0.001, 0.5176)$. The two types of soliton solutions can be seen in Figs.~\ref{orbits2}(d) and ~\ref{orbits3}(d), respectively, corresponding to branch 1 and 2 in Fig.~\ref{orbits1}(d).

\section{Analytical solution methods}
\label{AnalyticalSolutions}

\subsection{Oscillating multi-soliton solutions}
\label{Bsolitons}

\begin{figure}[t]
\centering
\subfigure{
\label{fig:ex3-a}
\includegraphics[width=.7\textwidth]{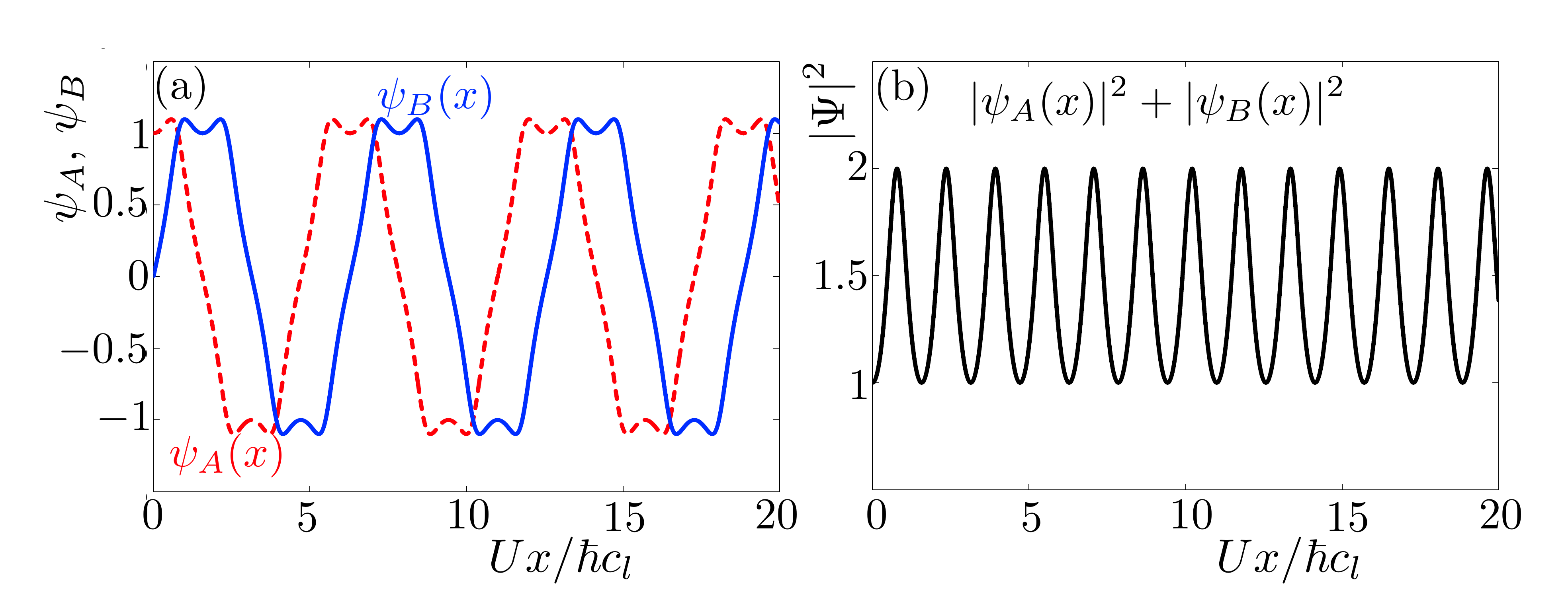}} 
\caption[]{\emph{Exact oscillating solution of the NLDE}. (a) The upper (red) and lower (blue) component solutions. (b) The spatial dependence of the total particle density. The spatial pattern of the density reveals the oscillating solution as a series of equally spaced bright solitons over a nonzero background.} \label{BrightSolitonTrain}
\end{figure}

Large amplitude oscillating solutions for strong nonlinearity such as those in Figs.~\ref{orbits2}(g)-(i) may be obtained analytically. Such solutions are bright solitons in the total density over a nonzero background. We note that analytical solutions in the massive case have recently been studied in detail by U. Al Khawaja~\cite{Alkhawaja2014}. We start by writing the two-spinor wavefunction in the form of a product of an envelope function $\eta(x)$ and the internal spinor degrees of freedom parameterized by the function $\varphi(x)$
\begin{eqnarray}
\Psi_\mathrm{zigzag}(x) = \eta(x) \left(\begin{array}{c c}
 \mathrm{cos}\varphi(x) \vspace{.3pc}\\
  \mathrm{sin}\varphi(x)   \label{basicform}
 \end{array} \right) \, , 
\end{eqnarray}
where we have assumed only that the wavefunction is real, i.e., choosing to work from the zigzag NLDE. Note that there is an arbitrary overall phase constant and translation symmetry $x \to x_0$, which may included in the final solution. Substituting Eq.~(\ref{basicform}) into Eqs.~(\ref{eqn:CondPsi9})-(\ref{eqn:CondPsi10}), multiplying by $\mathrm{cos}\varphi$ and  $\mathrm{sin}\varphi$, respectively, then adding the resulting equations gives
\begin{eqnarray}
\frac{d \varphi}{dx} = -\frac{\mu}{\hbar c_l} \left[ 1 - \eta^2 \, (U/\mu) ( \mathrm{cos}^4\varphi +  \mathrm{sin}^4\varphi) \right] . \label{phiprime}
\end{eqnarray}
To obtain a second equation we multiply Eqs.~(\ref{eqn:CondPsi9})-(\ref{eqn:CondPsi10}) by $\mathrm{cos}\varphi$ and $\mathrm{sin}\varphi$, respectively, then subtract the resulting equations which yields
\begin{eqnarray}
\frac{d(\mathrm{ln}\eta)}{dx} = \frac{U}{4 \hbar c_l}\, \eta^2 \, \mathrm{sin}(4 \varphi) \, . \label{etaprime}
\end{eqnarray}
Note that we have divided through by $\eta$ to arrive at Eqs.~(\ref{phiprime})-(\ref{etaprime}). Equations~(\ref{phiprime})-(\ref{etaprime}) can be combined by back substitution to get
\begin{eqnarray}
\left[ \frac{ \mathrm{sin}(4\varphi)} {\mathrm{cos}^4\varphi +  \mathrm{sin}^4\varphi} \right]\left(\varphi + \frac{\mu x}{\hbar c_l} \right)' = 4 \, (\mathrm{ln}\eta)'\,  \label{generalform}
\end{eqnarray}
where the prime notation indicates differentiation with respect to $x$. A formal expression for $\eta$ is obtained from Eq.~(\ref{generalform})
\begin{eqnarray}
\eta^4 = \mathrm{exp} \! \!\left[  \int \! dx \, \frac{ \mathrm{sin}(4\varphi)} {\mathrm{cos}^4\varphi +  \mathrm{sin}^4\varphi}  \left(\varphi + \frac{\mu x}{\hbar c_l} \right)'  \right] \, . \label{etaexpression}
\end{eqnarray}
To solve Eq.~(\ref{etaexpression}), we assume the linear form $\varphi(x) = \kappa x$, obtain an explicit form for $\eta(x)$, which we then substitute into Eq.~(\ref{phiprime}) to determine the constant $\kappa$ and obtain a relation for the chemical potential $\mu$ and the interaction $U$. Equation~(\ref{etaexpression}) becomes 
\begin{eqnarray}
\eta^4 = \mathrm{exp} \! \!\left[(\kappa + \mu /\hbar c_l )  \int \! dx \, \frac{ \mathrm{sin}(4 \kappa x)} {\mathrm{cos}^4 \kappa x +  \mathrm{sin}^4\kappa x}   \right] \, ,  \label{etaexpression2}
\end{eqnarray}
which, upon integration, yields the result
\begin{eqnarray}
\eta(x) = A  \left( \mathrm{cos}^4 \kappa x  + \mathrm{sin}^4 \kappa x  \right)^{-(1 + \mu/\kappa \hbar c_l )/4} , \label{etasolution2}
\end{eqnarray}
where $A$ is the integration constant. Substituting this result into Eq.~(\ref{phiprime}) and using the linear assumption, gives the expression
\begin{eqnarray}
\kappa = -\frac{\mu}{\hbar c_l} \left[ 1 - \frac{U}{\mu}  \left( \mathrm{cos}^4 \kappa x  + \mathrm{sin}^4 \kappa x  \right)^{ 1 -(1 + \mu/\kappa \hbar c_l )/2} \right] \, .  \label{kappacondition}
\end{eqnarray}
Since $\kappa$ is constant, it must be that the exponent of the spatial functions is identically zero. Equation~(\ref{kappacondition}) then gives the two conditions 
\begin{eqnarray}
\frac{1}{2} - \frac{\mu}{2 \kappa \hbar c_l}   =   0 \, , \\
- \frac{\mu}{\hbar c_l} \left( 1- \frac{U}{\mu} \right)  =   \kappa \, , 
\end{eqnarray}
which may be solved to give 
\begin{eqnarray}
\kappa  =  \frac{\mu}{\hbar c_l} \,, \\
\mu  =  2 \, U \, . 
\end{eqnarray}
The corresponding solution is then 
\begin{eqnarray}
\Psi(x) = A  \left( \mathrm{cos}^4\kappa x + \mathrm{sin}^4\kappa x \right)^{-1/2}  \left( \begin{array}{c} 
                                          \mathrm{cos} \, \kappa x     \\
                                         \vspace{-.7pc}    \\
                                             \mathrm{sin} \, \kappa x   \end{array}      \right)        \, ,    \label{brightsolitontrain}                      
\end{eqnarray}
where $\kappa = 2U/(\hbar c_l)$. The spinor components in Eq.~(\ref{brightsolitontrain}) are plotted in Fig.~\ref{BrightSolitonTrain}(a) and the corresponding density in Fig.~\ref{BrightSolitonTrain}(b). Although Eq.~(\ref{brightsolitontrain}) was obtained for the NLDE with real coefficients, a direct transformation to get the associated solution for the case of complex coefficients is obtained by taking $\psi_A \to i \psi_B$, $\psi_B \to  \psi_A$ to get 
\begin{eqnarray}
 \Psi(x) = \!  A\left( \mathrm{cos}^4\kappa x + \mathrm{sin}^4\kappa x \right)^{-1/2} \!  \left( \begin{array}{c} 
                                         i  \, \mathrm{sin} \, \kappa x     \\
                                         \vspace{-.7pc}    \\
                                             \mathrm{cos} \, \kappa x   \end{array}      \right)      .   \label{brightsolitontrain2}                      
\end{eqnarray}

A distinguishing feature of the solitons in Fig.~\ref{BrightSolitonTrain} is that the spinor component functions oscillate about zero but the total density does not exhibit nodes. From the density standpoint, the peaks in Fig.~\ref{BrightSolitonTrain}(b) are bright solitons on a nonzero background. 

\subsection{Dark and bright solitons by parametric transformation}
\label{ParametricSolitons}

In this section we isolate single dark and bright soliton solutions analytically using a parametric transformation motivated by the form of the invariance relation Eq.~(\ref{firstconservation}). A preliminary step requires that we solve Eq.~(\ref{firstconservation}) for $f_B$ and then back substitute into Eq.~(\ref{zigzag2}). This gives us the first order nonlinear equation
\begin{eqnarray}
 \frac{df_A}{dx}  = \mp  \frac{U}{\hbar c_l} \left\{ \left[  \frac{ \mu}{U}  \pm \sqrt{  2\frac{T^{11}}{U} + 2\left(  \frac{ \mu}{U} \right)^2 -   \left(f_A^2 -  \frac{\mu}{U} \right)^2} \right] \right. \\
\left.  \times \! \! \left[  2\frac{T^{11}}{U} + 2\left( \frac{ \mu}{U} \right)^2 -   \left(f_A^2 -  \frac{\mu}{U} \right)^2\right] \right\}  \! \!.  \label{fAprimeReduced}
\end{eqnarray} 
Equation~(\ref{fAprimeReduced}) is separable in the variables $f_A$ and $x$. In the special case where $\mu =0$, the problem simplifies and yields the integral
\begin{eqnarray}
\int \! df_A   \left( 2 \frac{T^{11}}{U} - f_A^4 \right)^{-3/4} =   \pm \frac{U}{  \hbar c_l} x + D \,  , \label{fAprime}
\end{eqnarray}
where $D$ is an integration constant. For $|f_A| < \left( 2T^{11} /U \right)^{1/4}$ and $T^{11}  > 0$, Eq.~(\ref{fAprime}) shows that $f_A^\prime$ will remain positive for all $x$; thus a monotonically increasing and bounded solution exists between $-  \left( 2T^{11} /U\right)^{1/4}     \!  < f_A <  \! + \left( 2T^{11} /U\right)^{1/4}$, for $-\infty < x< + \infty$. The integral in Eq.~(\ref{fAprime}) is given in terms of the hypergeometric function so that 
\begin{eqnarray}
\frac{U}{\hbar c_l} x  = \frac{f_A}{  \left( T^{11}/U \right)^{3/4}  } \;  _2F_1 \! \! \left( \frac{1}{4},\,  \frac{3}{4};\,  \frac{5}{4} ; \,  \frac{ U f_A^4}{  T^{11}      } \right) + D \,  . 
\end{eqnarray}
The second spinor component $f_B$ is obtained by back substitution into the invariance relation where we find $f_B = \left( 2T^{11} /U - f_A^4 \right)^{1/4}$. This type of solution exhibits a form of self-confinement similar to that studied in~\cite{santos2009}, where $f_B$ remains localized near the region where the slope of $f_A$ is steepest and has a shape similar to a $\mathrm{tanh}$ function.

\begin{figure}[t]
\centering
\subfigure{
\label{fig:ex3-a}
\includegraphics[width=.55\textwidth]{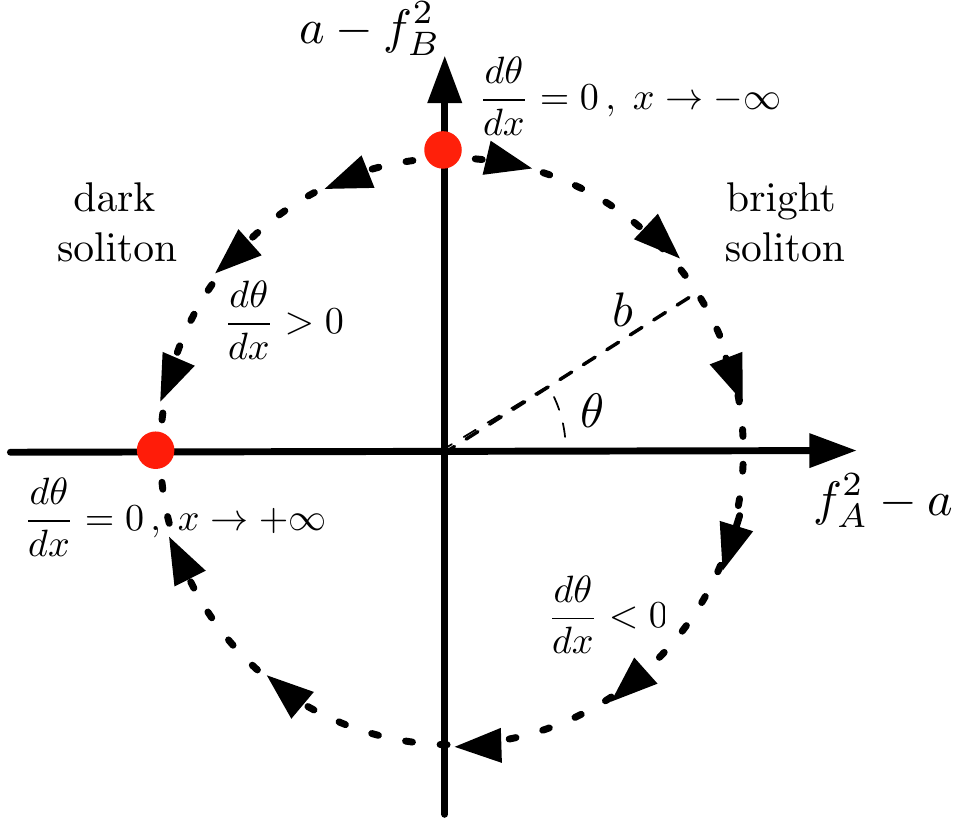}} \\
\caption[]{\emph{Graphical representation of dark and bright solitons using the parametric transformation}. Cast in terms of the angular parameter $\theta$, the dark soliton emerges as an interpolation between $\theta = \pi/2$ and $\theta = \pi$, in contrast to the bright soliton which interpolates between $\theta = \pi/2$ and $\theta = - \pi$.}
\label{Theta}
\end{figure}

For general values of $\mu$, it is helpful to cast Eq.~(\ref{fAprimeReduced}) in a more enlightening form by an appropriate parametric transformation. First, we make the substitution $f_A^2 =  a +  b \, \mathrm{cos}\theta $, with $a \equiv \mu/U$ and $b^2 \equiv 2T^{11}/U  + 2(\mu/U)^2$, which transforms Eq.~(\ref{fAprimeReduced}) to
\begin{eqnarray}
\frac{d\theta}{ dx} = \pm \,\frac{U}{\hbar c_l}   \sqrt{  a +  b\,  \mathrm{cos} \theta } \; \sqrt{ a \pm b\,  \mathrm{sin} \theta } \, . \label{ThetaEqn}
\end{eqnarray}
Note that in order to keep $f_A$ real we must enforce the constraint $a > b$. Cast in terms of the new variable $\theta$, we have the symmetric result $b \, \mathrm{cos}\theta =  f_A^2 - a$ and $b \, \mathrm{sin}\theta = a - f_B^2  $. Equation~(\ref{ThetaEqn}) has an interesting graphical representation which we have depicted in Fig.~\ref{Theta} where we show how the dark and bright solitons emerge when the problem is cast in terms of the angular parameter $\theta$.

The dark and bright solitons may be found in the limit that $a \to 1$, $b \to 1$, and with the choice $T^{11}= \mu^2/2U$ as in Sec.~\ref{GeneralSolutions}. In this case $d\theta/dx = 0$ at $\theta = \pi/2$ and $\theta = \pi$ for the negative sign under the radical in Eq.~(\ref{ThetaEqn}), and $d\theta/dx = 0$ at $\theta = - \pi/2$ and $\theta = \pi$ for the positive sign under the radical. In particular, for the negative sign under the radical, the dark soliton and bright soliton interpolate between $\theta = \pi/2$ and $\theta = \pi$ by following anti-clockwise and clockwise paths, respectively, i.e., for the positive and negative outer signs in Eq.~(\ref{ThetaEqn}).

\begin{figure}[h]
\centering
\subfigure{
\label{fig:ex3-a}
\includegraphics[width=.7\textwidth]{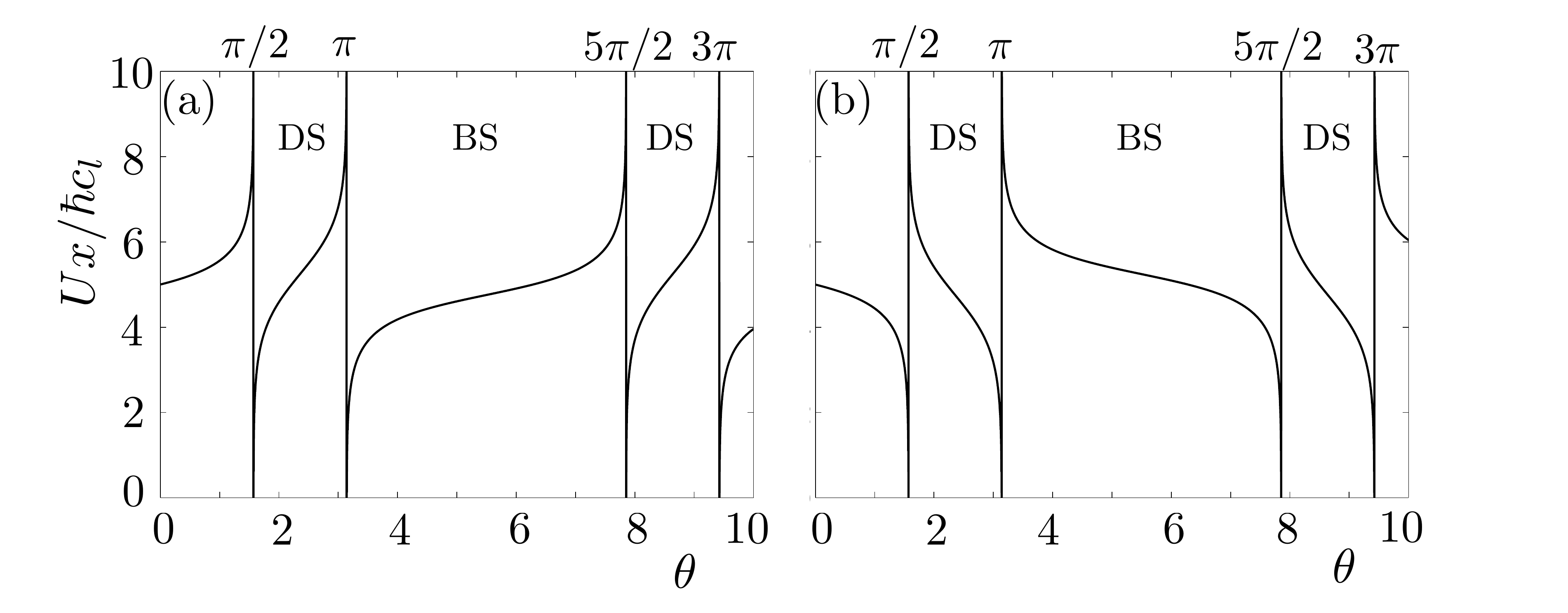}} \\ 
\caption[]{\emph{Solution of the parametric integral}. Solutions for the parameter $x(\theta)$ in Eq.~(\ref{xoftheta}) for (a) the positive sign (clockwise direction) and (b) the negative sign (anti-clockwise direction). To indicate the corresponding soliton solution, we use the shorthand DS = dark soliton, BS = bright soliton. The integration constant represents a spatial shift which we have taken as $D=0$. }
\label{ThetaPosNeg}
\end{figure}

Equation~(\ref{ThetaEqn}) may be integrated exactly by separation of variables whereby one obtains
\begin{eqnarray}
 \frac{U}{\hbar c_l} x + D =  \pm \, \frac{ \left( \mathrm{cos}\theta - \mathrm{sin}\theta + 1 \right) \mathrm{ln}\left[\left(  \mathrm{cos} \theta/2 \right) \left( \mathrm{cos}   \theta/2 - \mathrm{sin} \theta/2    \right)^{-1} \right]   }{\sqrt{ 1 + \mathrm{cos} \theta } \, \sqrt{ 1 - \mathrm{sin} \theta } } \, , \label{xoftheta} 
\end{eqnarray}
where $D$ is the integration constant which adds a spatial translation to the soliton core. Equation~(\ref{xoftheta}) cannot be directly inverted to find explicit forms for $f_A(x)$ and $f_B(x)$ (via $\theta(x)$), but we may study it graphically by plotting $U x/\hbar c_l$ as a function of $\theta$. In Fig.~\ref{ThetaPosNeg} we have plotted Eq.~(\ref{xoftheta}) where we have shown singular points occurring at repeating intervals: $..., \pi/2, \, \pi, \, 5\pi/2, \, 3\pi, \, ...$, alternating between the two types of soliton solutions.

\begin{figure}[t]
\centering
\subfigure{
\label{fig:ex3-a}
\includegraphics[width=.6\textwidth]{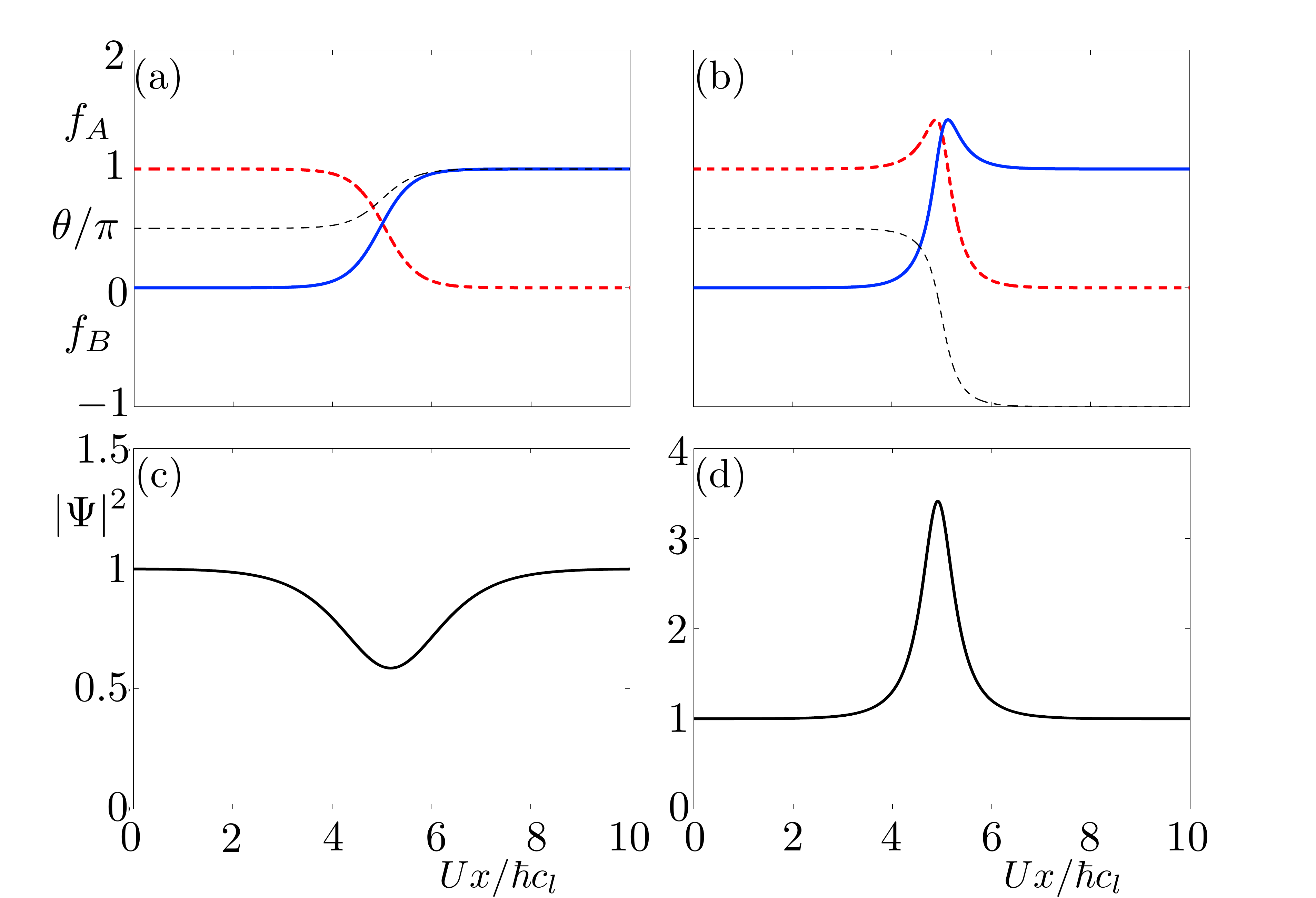}} \\ 
\caption[]{\emph{Soliton solutions obtained by the parametric method}. (a) The dark soliton and (b) bright soliton. The parametric function $\theta(x)$(dashed black) as shown in (a) and (b) matches the results in Fig.~\ref{ThetaPosNeg}. The spinor component solutions $f_A(x)$ (dashed red) and $f_B(x)$ (solid blue) are also shown here. }
\label{ThetaSolutions}
\end{figure}

Alternatively, Eq.~(\ref{ThetaEqn}) may be solved numerically to obtain an inversion of the solution plotted in Fig.~\ref{ThetaPosNeg}, i.e., $\theta(x)$. Figure~\ref{ThetaSolutions} shows this numerical result for the dark soliton in (a) and bright soliton in (b) with the densities for each soliton type shown in the lower panels. Note that the spinor components are nonzero within the soliton cores. Also, there is a clear signature associated with the densities of each soliton type: the dark soliton density dips to a factor of $0.6$ of the asymptotic background density whereas the bright soliton peaks at $3.5$ times the background.

The invariant form of Eq.~(\ref{conservedcharge}) says that asymptotically flat soliton solutions must have the same topological invariant quantity $\delta_{\pm \infty}$ given by Eq.~(\ref{conservedcharge2}). We verify that this is the case by explicitly computing $\delta_{\pm \infty}$ using Eq.~(\ref{ThetaEqn}) and expressing the integral in Eq.~(\ref{conservedcharge2}) in terms of the variable $\theta$. The resulting indefinite integral for $\delta_{\pm \infty}$ is 
\begin{eqnarray}
\hspace{0pc} \delta_{\pm \infty}  &= \int_{\theta_i}^{\theta_f} \! \! d\theta \sqrt{1 + \mathrm{cos} \theta} \, \sqrt{1 - \mathrm{sin} \theta}  \left( \mathrm{cos} \theta + \mathrm{sin}\theta \right) \label{Qintegral} \\
\hspace{0pc}  &= \frac{ \sqrt{ 1 + \mathrm{cos} \theta } \, \sqrt{ 1 - \mathrm{sin} \theta } \left[   \mathrm{sin} \theta+ \left(  \mathrm{sin}\theta - 1 \right) \mathrm{cos} \theta \right] }{ 1 + \mathrm{cos} \theta - \mathrm{sin}\theta } \left. \right\vert_{\theta_i}^{\theta_f} \nonumber . 
\end{eqnarray}
When evaluated at the appropriate limits for either of the two types of solitons, the integral in Eq.~(\ref{Qintegral}) yields $\delta_{\pm \infty}= 0$. For the dark soliton $\theta_i = \pi/2$ and $\theta_f = \pi$, whereas $\theta_i = \pi/2$ and $\theta_f = - \pi$ for the bright soliton. We point out that the dark and bright solitons are distinct and cannot be deformed into each other. In the case of the former, the crossing point where $f_A = f_B$ lies below the fixed point $(\sqrt{\mu/U}, \sqrt{\mu/U})$, whereas for the latter the crossing point lies above the fixed point. Any continuous deformation in the NLDE solution space which moves the crossing point towards the fixed point must force $f_A$ and $f_B$ to flatten out everywhere pushing the crossing point out to $x \to \infty$. This analysis demonstrates that our solitons are indeed of two distinct types, unrelated by any continuous transformation in parameter space. 

The character of the parametrization angle $\theta$ in Fig.~\ref{Theta} illustrates this point. For example, we are free to parametrize our problem more generally in terms of some variable $\tau$: $b \to b(\tau)$ and $\theta \to \theta(\tau)$ in Eq.~(\ref{ThetaEqn}). And, since the singular point $(b =0, \, \theta)$ is a fixed point of Eqs.~(\ref{zigzag1})-(\ref{zigzag2}), any continuous deformation of the functions $b(\tau)$ and $\theta(\tau)$ is allowed provided we avoid the origin where $b = 0$. Thus, just the fact that the dark and bright soliton paths encircle the origin in opposite directions suggests distinct homotopy classes for the two types of solitons.

It is important to emphasize that we classify our solitons as dark or bright based on the density profiles shown in Fig.~(\ref{ThetaSolutions})(c) and (d). For our single isolated solitons the densities occur as either a suppression (notch) or elevation (peak) with respect to a nonzero condensate background: no nodes are observed in the density profiles. Nevertheless, the underlying spinor components asymptotically approach zero in one direction. This is in marked contrast to the usual case of a single component BEC. For instance, bright solitons occur in presence of an attractive interaction, as in general NLSE based theories with focusing nonlinearity (attractive), while a defocusing nonlinearity (repulsive) leads to a dark soliton profile. It is worth noting that dark and bright solitons may occur in the same system such as in extended Bose-Hubbard models with strong repulsive on-site interactions~\cite{Balakrishnan2009,Reinhardt2011,Rubbo2012}. A key feature of this particular model is that the binary nature of the on-site occupation number allows for a mapping to a spin-1/2 system. The particle-hole asymmetry in the spin-1/2 model characterizes the crossover from a nonlinear Schr\"odinger order parameter to that of a strongly repulsive system and gives rise to dark and bright multi-species setting. In spite of these similarities, our solutions are multi-component bright and dark solitons directly relating to the relativistic context.

\subsection{Soliton series expansions}
\label{SolitonExpansion}

A fourth approach to obtain solutions of the NLDE is through a series expansion. This approach allows for general values of the ratio $\mu/U$, interpolating through the soliton boundary and connecting the two regimes shown in Fig.~\ref{orbits1}. The series expansion uses the same ansatz as that in Sec.~\ref{Bsolitons}, i.e., Eq.~(\ref{basicform}), whereby we find that $\eta(x)$ will take the form of a power series in the quantity $\mathrm{cos}^4\varphi + \mathrm{sin}^4\varphi$. The function $\varphi(x)$ can then be obtained by integrating $\eta(x)$ term by term. This offers a convenient method since powers of $\mathrm{cos}^4\varphi + \mathrm{sin}^4\varphi$ are exactly integrable.

We begin by substituting the form $\Psi(x) = \eta(x) \left( \mathrm{cos}\varphi(x) , \, \mathrm{sin}\varphi(x) \right)$ into the invariance relation Eq.~(\ref{firstconservation}). Upon this substitution, Eq.~(\ref{firstconservation}) becomes 
\begin{eqnarray}
\hspace{-1pc} \eta^2 =      \frac{ 1}{ \mathrm{cos}^4 \varphi + \mathrm{sin}^4 \varphi } \left( \frac{ \mu }{U} \right)  \! \! \left[ 1 \pm \sqrt{ 1 +  2\left( \frac{T^{11} }{U}\right)  \! \left( \frac{U}{\mu} \right)^2  ( \mathrm{cos}^4 \varphi + \mathrm{sin}^4 \varphi ) }  \right] ,    \label{conservedeta} 
\end{eqnarray}
and the associated equation for $\varphi(x)$ is Eq.~(\ref{phiprime}) which takes the form
\begin{eqnarray}
\frac{d \varphi }{ dx} = \pm \,  \frac{\mu}{\hbar c_l}   \sqrt{ 1 +   2\left( \frac{T^{11} }{U}\right) \left( \frac{U}{\mu} \right)^2  \!( \mathrm{cos}^4 \varphi + \mathrm{sin}^4 \varphi ) } \,  . \label{conservedphiprime}
\end{eqnarray}
Because of the bounds $ 1/2 \le \mathrm{cos}^4 \varphi + \mathrm{sin}^4 \varphi \le 1$, there are two limits for which Eqs.~(\ref{conservedeta})-(\ref{conservedphiprime}) simplify, defined by $| (2T^{11}/U) (U/ \sqrt{2} \mu)|^2 \gg 1$ and $| (2T^{11}/U) (U/\mu)|^2 \ll1$. For $| (2T^{11}/U) (U/ \sqrt{2} \mu)|^2 \gg 1$, Eqs.~(\ref{conservedeta})-(\ref{conservedphiprime}) may be expanded in an asymptotic series, whereby one finds
\begin{eqnarray}
  \eta(x)^2 = |2 T^{11}/U|^{1/2}  \left( \mathrm{cos}^4 \varphi(x)+ \mathrm{sin}^4 \varphi(x) \right)^{-1/2}  \\
             \times  \sum_{n=0}^{\infty} \left(\! \! \begin{array}{c}   1/2 \\ n \end{array} \!  \! \right) \left[  (2T^{11}/U) (U/\mu)^2  \left( \mathrm{cos}^4 \varphi(x)+ \mathrm{sin}^4 \varphi(x) \right) \right]^{-n} , \nonumber \\
\nonumber \\
 \nonumber \\
 \frac{d}{dx}  \varphi(x) =  \pm \frac{\mu}{\hbar c_l}  \left( \mathrm{cos}^4 \varphi(x)+ \mathrm{sin}^4 \varphi(x) \right)^{1/2}  \\
\times  \sum_{n=0}^{\infty} \left(\! \! \begin{array}{c}   1/2 \\ n \end{array} \!  \! \right) \left[ (2 T^{11}/U)(U/\mu)^2  \left( \mathrm{cos}^4 \varphi(x)+ \mathrm{sin}^4 \varphi(x) \right) \right]^{-n},   \nonumber 
\end{eqnarray}
where the expansion coefficients are the generalized binomial coefficients 
\begin{eqnarray}
 \left(\! \! \begin{array}{c}   1/2 \\ n \end{array} \!  \! \right) &\equiv  \frac{ 1/2 (1/2 -1 ) (1/2-2)  \cdots }{ n !  \,  (1/2- n) (1/2-n -1) (1/2-n -2) \cdots } \nonumber \\
 &=  \frac{ 1/2 (1/2 -1 )\cdots  (1/2-n + 1) }{ n !  }  \, .
 \end{eqnarray}
The terms in the expansion for $\varphi(x)$ can be integrated and expressed in terms of elliptic integrals. The leading order terms in the asymptotic expansion are
\begin{eqnarray}
\hspace{-3pc}  \eta(x) \approx | 2T^{11}/U|^{1/4}  \left( \mathrm{cos}^4 \varphi(x)+ \mathrm{sin}^4 \varphi(x) \right)^{-1/4} + \mathcal{O}\left[ (2 T^{11}/U) (U/ \sqrt{2} \mu)^2\right]^{-1} , \\
 \hspace{-3pc}     \nonumber \\
     \hspace{-3pc}     \varphi(x) \approx \pm  \mathrm{E}^{-1}\! \!  \left( \frac{2  T^{11}Ux}{\hbar c_l} \Bigg| \frac{1}{2} \right) + \mathcal{O}\left[ (2 T^{11}/U)(U/ \sqrt{2} \mu)^2\right]^{-1}  , 
\end{eqnarray}
where the solution for $\varphi(x)$ is expressed in terms of the inverse elliptic integral of the first kind~\cite{Stegun1972}. Conversely, at the other extreme limit $| (2 T^{11}/U)(U/\mu)|^2 \ll 1$ the following expansion is valid
\begin{eqnarray}
\hspace{-2pc}  \eta(x)^2 =  \left( \mathrm{cos}^4 \varphi(x)+ \mathrm{sin}^4 \varphi(x) \right)^{-1}\left( \frac{\mu}{U}\right) \label{etaexpansion} \\
          \times \left[  1 \pm \sum_{n=0}^{\infty} \left(\! \! \begin{array}{c}   1/2 \\ n \end{array} \!  \! \right) \left[  (2 T^{11}/U)(U/\mu)^2  \left( \mathrm{cos}^4 \varphi(x)+ \mathrm{sin}^4 \varphi(x) \right) \right]^{n}  \right] , \nonumber \\
\hspace{-2pc}  \nonumber \\
\hspace{-2pc}  \nonumber \\
 \hspace{-2pc}  \frac{d}{dx}  \varphi(x)=    \pm \frac{\mu}{\hbar c_l}  \sum_{n=0}^{\infty} \left(\! \! \begin{array}{c}   1/2 \\ n \end{array} \!  \! \right) \left[  (2 T^{11}/U)(U/\mu)^2  \left( \mathrm{cos}^4 \varphi(x)+ \mathrm{sin}^4 \varphi(x) \right) \right]^{n} \, .   
\end{eqnarray}
To leading order, only one nonzero solution exists, coming from the positive sign inside the square brackets in Eq.~(\ref{conservedeta}): 
\begin{eqnarray}
\eta(x) \approx  \frac{( U/2 \mu)^{1/2} }{ \left[ \mathrm{cos}^4 \varphi(x)+ \mathrm{sin}^4 \varphi(x) \right]^{1/2} }+ \mathcal{O}\left[(2 T^{11}/U)  (U/  \mu)^2\right] , \\
   \nonumber \\
 \varphi(x) \approx  \pm \frac{\mu x }{\hbar c_l}  + \mathcal{O}\left[ (2 T^{11}/U) (U/  \mu)^2\right]  . \label{originalsoln}
\end{eqnarray}
We point out that Eq.~(\ref{originalsoln}) is just the series of bright solitons first obtained in Eq.~(\ref{brightsolitontrain}) which we now identify as the large $\mu/U$ limit solution in Fig.~\ref{orbits2}(i). A second solution can be obtained associated with the negative sign inside the brackets in Eq.~(\ref{conservedeta}), coming from the next to leading order term in Eq.~(\ref{etaexpansion}):
\begin{eqnarray}
 \eta(x) \approx  \sqrt{\frac{| T^{11}| }{2 \mu} }   + \mathcal{O}\left[ (2 T^{11}/U)(U/  \mu)^2\right]^2 , \label{planewave1} \\ 
 \nonumber \\
 \varphi(x) \approx  \pm \frac{\mu x }{\hbar c_l}  + \mathcal{O}\left[  (2 T^{11}/U)(U/  \mu)^2\right]  \, . \label{planewave2} 
\end{eqnarray}
Equations~(\ref{planewave1})-(\ref{planewave2}) describe plane wave solutions identified with the oscillating solutions in panel Fig.~\ref{orbits3}(i). To make the connection to the single dark and bright solitons in Figs.~\ref{orbits2}(d) and \ref{orbits3}(d), we set $ T^{11}= \mu^2/2U$ then Taylor expand Eqs.~(\ref{conservedeta})-(\ref{conservedphiprime}) in powers of $| 1 - (U/\mu)|$ for $\mu/U \approx 1$. Such an expansion is valid as long as $| 1 - (U/\mu)| < 1$. Solving Eqs.~(\ref{conservedeta})-(\ref{conservedphiprime}) to leading order in $| 1 - (U/\mu)|$ gives
\begin{eqnarray}
 \eta^2(x)  \approx   \frac{ 1}{ \mathrm{cos}^4 \varphi(x) + \mathrm{sin}^4 \varphi(x) } \\
 \times  \left[ 1 \pm \sqrt{ 1 - ( \mathrm{cos}^4 \varphi(x) + \mathrm{sin}^4 \varphi(x) ) }  \right] + \mathcal{O}\left[ | 1 - (U/\mu)| \right]  ,   \nonumber \label{darkbrighteta}\\
 \nonumber \\
   \frac{d }{dx}\varphi(x) \approx  \pm  \frac{\mu}{\hbar c_l}   \sqrt{ 1 - ( \mathrm{cos}^4 \varphi(x) + \mathrm{sin}^4 \varphi(x) ) } +  \mathcal{O}\left[ | 1 - (U/\mu)| \right]   .
  \end{eqnarray}
A solution for $\varphi(x)$ can be obtained by the method of separation of variables, whereby we find
\begin{eqnarray}
 \frac{ \mathrm{sin}(2 \varphi ) \,  \mathrm{ln} \left[ \mathrm{tan}(2 \varphi )   \right]     }{ \sqrt{ 1 - \mathrm{cos}(4 \varphi )   }} = \pm  \frac{\mu}{\hbar c_l}  x + D \, , \label{darkbrightphi}
\end{eqnarray}
where $D$ is the integration constant which spatially shifts the solution, as found also in Eqs.~(\ref{fAprime})-(\ref{xoftheta}). One may invert Eq.~(\ref{darkbrightphi}) graphically and substitute this result into Eq.~(\ref{darkbrighteta}), leading to the dark and bright solitons, respectively, for the positive and negative signs in Eq.~(\ref{darkbrighteta}).

\section{Solitons by the method of numerical shooting}
\label{NumericalSolitons}

So far we have focused our attention on solving the NLDE using analytical methods. In Sec.~\ref{NumericalSolitons}, we provide a detailed analysis of solutions by the method of numerical shooting. Numerics provide a versatile angle of attack and prove especially convenient for solitons confined by an external trap. Our numerical approach in this section is similar to that used for studying trapped BECs in the absence of a lattice background~\cite{Carr2006}.

\begin{figure}[t]
\centering
 \subfigure{
\label{fig:ex3-a}
\includegraphics[width=.65\textwidth]{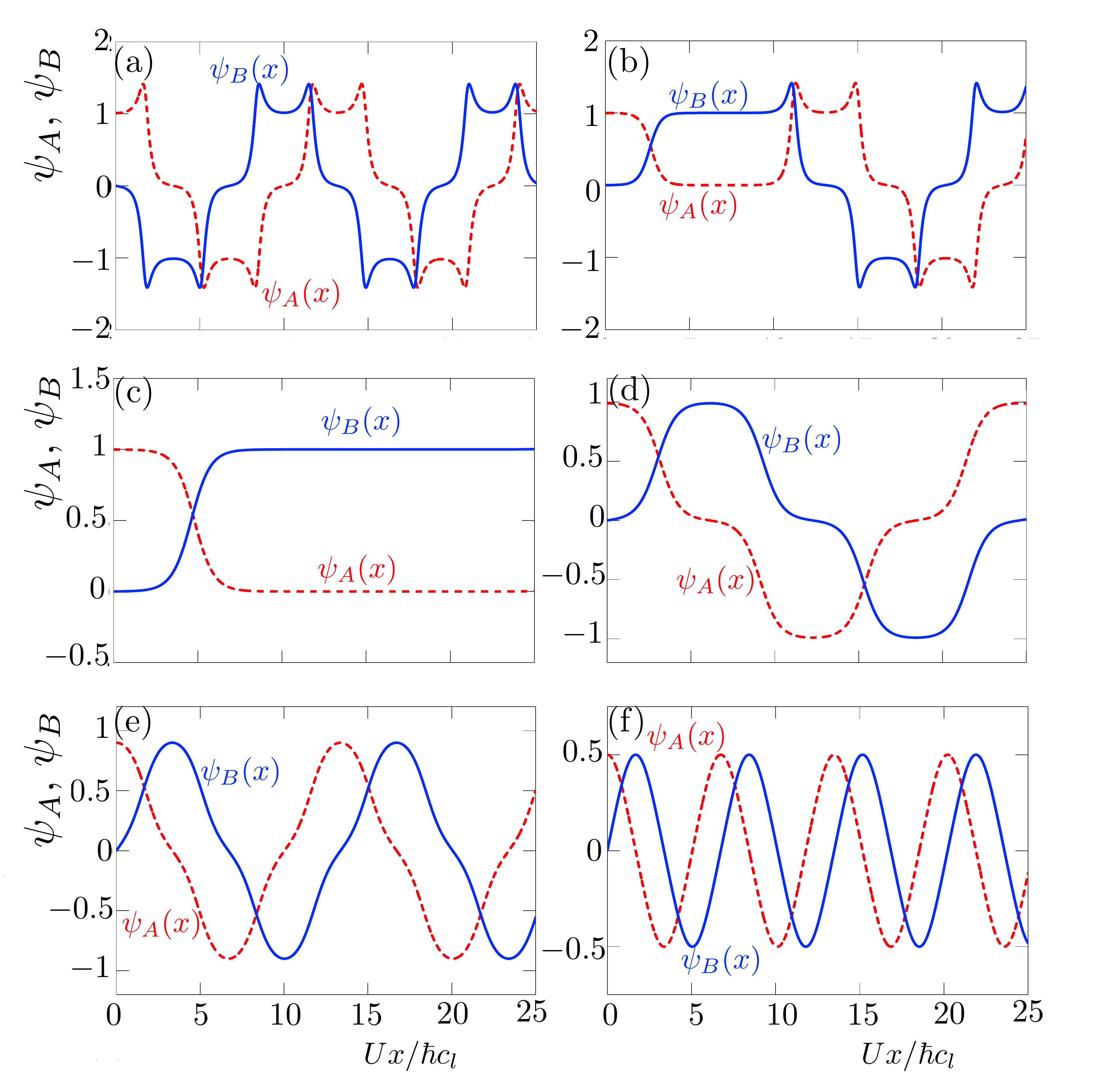}}\\ 
\caption[]{\emph{Numerical solutions of the NLDE}. (a)-(b) Overshooting the dark soliton solution. Here the initial value for $a_0$ is slightly larger than $a_0^\mathrm{soliton}$ and the oscillations are in the strong nonlinear regime. (c) The dark soliton appears as the oscillations are pushed away from $x=0$ by fine tuning $a_0$. (d)-(f) Undershooting the dark soliton. In this case $a_0$ is smaller than $a_0^\mathrm{soliton}$ with the solution approaching the linear plane-wave regime.} \label{ShootingSoliton}
\end{figure}

The most direct approach is to express Eqs.~(\ref{eqn:CondPsi9})-(\ref{eqn:CondPsi10}) in terms of the dimensionless spatial variable $\chi \equiv (U/\hbar c_l)x$. The spinor functions $f_A(\chi)$ and $f_B(\chi)$ are then expanded in a power series around $\chi=0$
\begin{eqnarray}
f_A(\chi) = \sum_{j=0}^\infty a_j \chi^j \, , \;\;\;\;  f_B(\chi) = \sum_{j=0}^\infty b_j \chi^j \, , \label{expansions}
\end{eqnarray}
with $a_j$ and $b_j$ the expansion coefficients to be determined. Since we are solving two coupled first order equations, we require the initial conditions $f_A(0)$ and $f_B(0)$. Substituting Eq.~(\ref{expansions}) into Eqs.~(\ref{eqn:CondPsi9})-(\ref{eqn:CondPsi10}) gives us the behavior of the solution at the origin
\begin{eqnarray}
f_A'(0) \sim \frac{1}{\hbar c_l} \left[\mu - U f_B(0)^2 \right] f_B(0)  \, ,    \\
f_B'(0) \sim -  \frac{1}{\hbar c_l} \left[\mu - U f_A(0)^2 \right] f_A(0). \label{corevalues}
\end{eqnarray}
By examination we find that for $j \ge 0$ the NLDE implies a recursion relation for the expansion coefficients $a_{j}$ and $b_{j}$. Recursion relations for the lowest values of the index $j$ are  
\begin{eqnarray}
    b_1 - a_0^3 = - a_0              \,  ,    \\
      2 b_2 -3 a_1 a_0^2 =  - a_1    \, ,    \\
     3 b_3  - 3 a_2 a_0^2 - 3 a_0 a_1^2 =  -  a_2    \, , \\
     4  b_4 - 3 a_3 a_0^2- 3 a_2 a_1^2-  a_1^3    =  - a_3    \,  ,   
\end{eqnarray}
resulting from Eq.~(\ref{eqn:CondPsi9}), and 
\begin{eqnarray}
  -  a_1- b_0^3  = -  b_0  \, , \\
   -  2 a_2 - 3 b_1 b_0^2 = -  b_1 \, , \\
   -  3   a_3   - 3 b_2 b_0^2 - 3 b_0 b_1^2= - b_2 \, ,  \\
  -  4  a_4 - 3 b_3 b_0^2- 3 b_2 b_1^2-  b_1^3   = -  b_3  \, , 
\end{eqnarray}
which come from Eq.~(\ref{eqn:CondPsi10}). To obtain soliton solutions using the shooting method, we first fix either $a_0$ or $b_0$, and then vary the other until we obtain convergence to the desired precision. Although it seems that both $a_0$ and $b_0$ are free parameters, fixing one to a different value before shooting results in a spatially translated final solution, as we verified numerically. This is true as long as the first parameter is fixed to a value between zero and one, for the dark soliton, or between one and the value of the peak for the bright soliton. The second ``shooting'' parameter can then be tuned to find the stable soliton solution. Taking $b_0 = 0$ and iterating to obtain the value for $a_0 = a_0^{\mathrm{soliton}}$ to the desired precision gives the soliton configuration. Figure~\ref{ShootingSoliton} shows the shooting process as we tune $a_0$ through the dark soliton which is shown in Fig.~\ref{ShootingSoliton}(c) for which $a_0^{\mathrm{soliton}}= 0.999292150145 \pm 10^{-8}$. Note that converging to a value $a_0^{\mathrm{soliton}} < 1$ ensures that the total density remains in the dark soliton regime. A similar process converges on a second value $a_0^{\mathrm{soliton}} > 1$ associated with the bright soliton.

As the precision in the value of $a_0^{\mathrm{soliton}}$ is increased, oscillations are pushed out to larger values of $x$. In Figs.~\ref{ShootingSoliton}(a) and (b) we show the solution for values of $a_0 > a_0^{\mathrm{soliton}}$. At larger values of $a_0$, the nonlinearity becomes dominant and we start to pick up some of the excited soliton states which can be seen in Figs.~\ref{ShootingSoliton}(a) and (b). For values $a_0 < a_0^{\mathrm{soliton}}$, the effect of the interaction is reduced, Figs.~\ref{ShootingSoliton}(d) and (e), until finally we see the free particle sine and cosine forms appearing in Fig.~\ref{ShootingSoliton}(f). The particular values of the constant $a_0$ in Fig.~\ref{ShootingSoliton} are: (a) $a_0 = 1.1\pm 10^{-19}$, (b) $a_0 = 0.9992922 \pm 10^{-13}$, (c) $a_0 = 0.999292150145 \pm 10^{-8}$, (d) $a_0 = 0.99 \pm 10^{-18}$, (e) $a_0 = 0.9\pm 10^{-19}$, (f) $a_0 =0.5 \pm 10^{-19}$. The bright soliton can be obtained using a similar shooting process. A similar method has been used to study convergent ring dark and bright solitons~\cite{CarCla06}.

\section{Conclusion} 
\label{Conclusion}

 In this article we have presented a variety of methods for solving the armchair nonlinear Dirac equation (NLDE) and mapped out the soliton landscape. A discrete symmetry allows for  two types of NLDEs associated with the quasi-one-dimensional reduction to the armchair geometry in the plane of a honeycomb optical lattice. This discrete symmetry is expressed in terms of Pauli matrices acting on the order parameter for bosons propagating along the length of the armchair pattern in the lattice.

 In particular, we have found dark and bright solitons for nonzero chemical potential and a self-confined soliton for the case $\mu=0$. Both dark and bright solitons show high-contrast density fringes and are therefore clearly observable in experiments. The spinor component functions for the dark soliton are approximated by the forms $\psi_A(x) \sim (1/2)[1 + \mathrm{tanh}(x)]$ and $\psi_B(x) \sim (1/2)[1 - \mathrm{tanh}(x)]$, while the bright soliton  components resemble the forms $\psi_A(x) \sim  (1/2)[1 + \mathrm{tanh}(x) + \mathrm{sech}(x) ]$ and $\psi_B(x) \sim  (1/2)[1 - \mathrm{tanh}(x)  + \mathrm{sech}(x) ]$, where $\psi_A(x)$ and $\psi_B(x)$ are the spatial parts of the upper and lower two-spinor components. The crossing point where $\psi_A = \psi_B$ lies below the fixed point $\psi_A = \psi_B = \sqrt{\mu/U}$, for the dark soliton, and above the fixed point in the case of the bright soliton. We have found that a continuous deformation between the dark soliton and bright soliton forces the solution to flatten out everywhere when pushing through the fixed point. Thus, the two solution regimes, large and small local densities, are topologically distinct. In the case where $\mu=0$, a third soliton solution exists which resembles a flattened $\mathrm{sech}$ function in $\psi_B$ localized at the center of a $\mathrm{tanh}$ form in $\psi_A$.

A distinguishing feature of our results is that our dark soliton does not exhibit a node in the total density, even though there are asymptotic zeros in the individual spinor components. An analogous effect occurs in dark solitons in Fermi gases, where a node appears in the gap function $\Delta(x)$, but not the total density. In the case of the Fermi gas, a dark soliton may be observed in the gap function $\Delta(x)$, the order parameter that encodes pairing of fermion particles and holes in the Bardeen-Cooper-Schrieffer (BCS) to Bose-Einstein condensate crossover. There the notch depth in $\Delta(x)$ varies with the translational speed of the soliton. The speed is characterized by an overall phase $\phi(x)$ which itself varies across the gap notch. Moreover, the density of Bogoliubov modes  inherits a similar gray soliton profile and has been investigated in recent theoretical and experimental work~\cite{Spuntarelli2011,Zwierlein2013}. The soliton notch in both the gap and the density becomes deepest and narrowest at unitarity: the regime between BCS and BEC where the pairing length is on the order of the atomic spacing.

Similarities to the Fermi gas brings up several questions worth addressing, suggesting possible future research topics. First, it is interesting to consider how the combined tuning of a complex gap and atom-atom interactions in the NLDE affect the notch depth and overall phase through the core of our solitons. Second, does a finite velocity boost induce Friedel-like oscillations in NLDE solitons similar to those studied in~\cite{Spuntarelli2011}? All such questions relate to the issue of Cooper pairing between NLDE quasi-particles, the relationship between composite particle-hole pairs and the fundamental lattice bosons, and the role of the NLDE in unitary Fermi gas analogs in general.

\ack{This material is based in part upon work supported by the National Science Foundation under grant number PHY-1067973. L.D.C. thanks the Alexander von Humboldt foundation and the Heidelberg Center for Quantum Dynamics for additional support. We acknowledge useful discussions with Ken O'Hara at Pennsylvania State University. }

\appendix 

\section{Convergence of numerical solutions of the quasi-1D reduction of the NLDE}

\begin{figure}[h]
\centering
\subfigure{
\label{fig:ex3-a}
\includegraphics[width=\textwidth]{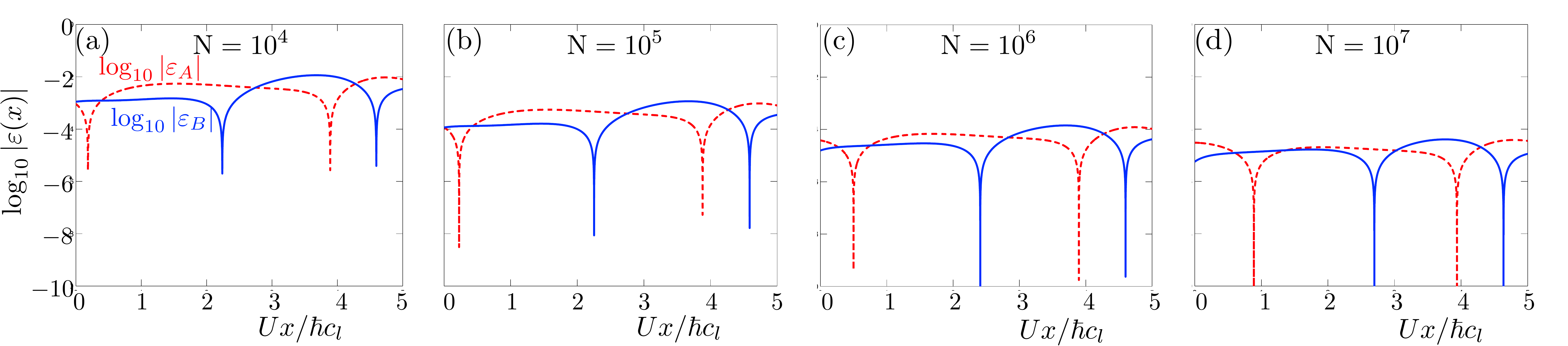}}\\ 
\caption[]{\emph{Convergence of oscillating solution of the quasi-1D reduction of the NLDE}. Error in the periodicity of the solution in Fig.~\ref{orbits2}(a). Solutions are obtained by finite differencing with the boundary values $f_A(0)$ and $f_B(0)$ obtained through the invariance relation. } \label{Convergence1}
\end{figure}

To compute the error of our oscillating numerical solutions, we first calculate the average difference between values of $\psi_A$ and $\psi_B$ at positions separated by one period. This tells us how the error in the periodicity of our solutions propagates with increasing position. The formula we use for the error $\varepsilon(x)$ is 
\begin{eqnarray}
\hspace{-1pc} \varepsilon_{A(B)}(x)  &\equiv& \frac{ \mathrm{diff}(x) }{ \mathrm{avg}(x) } =  2\left[   \frac{ \psi_{A(B)}( x + L )  - \psi_{A(B)}(x)}{ \psi_{A(B)}( x + L )  + \psi_{A(B)}(x)} \right]  \, , \label{periodicity}
\end{eqnarray}
where $L$ is the periodicity for the particular solution and the error is computed for both two-spinor component functions $\psi_A$ and $\psi_B$. In Fig.~\ref{Convergence1}, we have plotted $\mathrm{log}_{10}\left| \varepsilon(x) \right|$, (a)-(d), and $\varepsilon(x)$, (e)-(h), for the solution depicted in Fig.~\ref{orbits2}(a) using grid sizes $\mathrm{N} = 10^4, \, 10^5, \, 10^6, \, 10^7$. The solution in Fig.~\ref{orbits2}(a) is obtained by using a forward stepping finite difference scheme. The initial conditions are obtained by choosing a value for $f_A$ which lies on the desired solution branch then substituting into the invariance relation Eq.~(\ref{fBinitialcond}) to obtain $f_B(0)$.

\section*{References}



\bibliographystyle{unsrt}

\bibliography{NLDE_Solitons_Refs}


\end{document}